\documentclass[amsmath,amssymb,aps,pre,reprint]{revtex4-1}

\usepackage{graphicx}
\usepackage{dcolumn}
\usepackage{bm}
\usepackage{color}
\usepackage{dsfont}

\newcommand{\eqn}[1]{\begin{eqnarray}#1\end{eqnarray}}

\newcommand{\ave}[1]{\left\langle #1 \right\rangle}
\newcommand{\kb}[0]{k_{\rm B}}
\newcommand{\dd}[0]{\partial}
\newcommand{\mC}[0]{\mathcal C}
\newcommand{\mF}[0]{\mathcal F}
\newcommand{\mG}[0]{\mathcal G}
\newcommand{\mO}[0]{\mathcal O}
\newcommand{\mR}[0]{\mathcal R}
\newcommand{\fR}[0]{\mathfrak R}
\newcommand{\fG}[0]{\mathfrak G}
\newcommand{\fA}[0]{\mathfrak A}
\newcommand{\fB}[0]{\mathfrak B}
\newcommand{\mL}[0]{\mathcal L}
\newcommand{\fL}[0]{\mathfrak L}
\newcommand{\fK}[0]{\mathfrak K}
\newcommand{\fP}[0]{\mathfrak P}
\newcommand{\Lam}[0]{\{\Lambda^\nu\}}
\newcommand{\od}[0]{^{\rm od}}
\newcommand{\ef}[0]{^{\rm ef}}

\begin{document}

\title{Overdamped stochastic thermodynamics with multiple reservoirs}

\author{Y\^uto Murashita}
\affiliation{Department of Physics, University of Tokyo, 7-3-1 Hongo, Bunkyo-ku, Tokyo, 113-0033, Japan}
\author{Massimiliano Esposito}
\affiliation{Complex Systems and Statistical Mechanics, Physics and Materials Science, University of Luxembourg, L-1511 Luxembourg, Luxembourg}

\date{\today}

\begin{abstract}
After establishing stochastic thermodynamics for underdamped Langevin systems in contact with multiple reservoirs,
we derive its overdamped limit using timescale separation techniques. The overdamped theory is different from the 
naive theory that one obtains when starting from overdamped Langevin or Fokker-Planck dynamics and only coincide with it in presence 
of a single reservoir. The reason is that the coarse-grained fast momenta dynamics reaches a nonequilibrium state 
which conducts heat in presence of multiple reservoirs. The underdamped and overdamped theory are both shown to 
satisfy fundamental fluctuation theorems. Their predictions for the heat statistics are derived analytically for a Brownian 
particle on a ring in contact with two reservoirs and subjected to a non-conservative force and are shown to coincide 
in the long-time limit.
\end{abstract}

\maketitle

\section{Introduction}

Langevin equations provide a simple description of physical, chemical and economical phenomena (see, e.g., Ref.~\cite{Gardiner}).
Originally, motivated by thermodynamic considerations, Langevin proposed an equation containing a stochastic noise to describe the dynamics of a diffusing Brownian particle and reproduce the Einstein relation in a simpler way \cite{Langevin1908, LG97}.
More recently, Sekimoto defined heat along each stochastic solution of the Langevin equation 
and by doing so endowed the Langevin dynamics with a thermodynamic interpretation \cite{Sek97, Sekimoto}.
This development, together with the discovery of fluctuation theorems \cite{JarzynskiRev11, Sei12, EspVDBRev2014} and the technological advances in the manipulation of small systems \cite{WSMe02,LDSe02,CRJe05,TSUe10,GVNe11,BAPe12,Bechhoefer2014}, gave rise to a new thermodynamics theory for small systems, nowadays called stochastic thermodynamics.
The overdamped Langevin equation and the Fokker-Planck equation, which is dynamically equivalent to the overdamped Langevin equation, have played crucial roles to delve into stochastic thermodynamics of isothermal systems \cite{Sei12}.

The problem of diffusion in presence of non-homogeneous temperature has a long history which starts with the theoretical studies of Landauer \cite{Lan75, Lan88}, B\"uttiker \cite{ButtikerZP98} and van Kampen \cite{vK87,vK88}.
Recently diffusion experiments with non-uniform temperature have been realized \cite{DB06, PMLe14} and created a renewed interest for theoretical studies in this field \cite{MatsuoSasa00, Kroy2010, CBEA12, Pol13, BC14, San15}. It has been demonstrated that a standard overdamped Langevin description fails to correctly evaluate thermodynamic quantities such as the 
entropy production, and that consequently one has to start from an underdamped description and construct a non-trivial overdamped approximation \cite{MatsuoSasa00, CBEA12, BC14}.

In this paper, we consider a different class of non-isothermal systems where the system is simultaneously coupled to multiple heat reservoirs with different temperatures.
These systems are crucial to build models of Brownian heat engines and they can be experimentally realized. 
They have been used for instance to experimentally verify fluctuation theorems for heat transfers \cite{GSBPC10, CINT13, CilibertoImparatoJSM13}. 
They could also be used to study efficiency fluctuations of heat engines which are nowadays actively studied \cite{VEWB14, VWvBE14, PVE15}.

Since directly calculating heat statistics analytically for underdamped dynamics is in many cases despairingly difficult (see Refs.~\cite{Vis06, Sab12, FogedbyImparato12} for exceptions), we would like to establish a much simpler overdamped description of the system.
The overdamped description is known to provide a correct dynamical (i.e. probability density in configuration space) 
and thermodynamical (i.e. statistics of heat flows) description in presence of a single reservoir. 
However, its naive extension to situations with multiple reservoirs, although properly describing the dynamics, dramatically fails to properly evaluate the thermodynamics.
If we use the overdamped Langevin equation, the heat flows between reservoirs coupled by one degree of freedom diverge (e.g. models of Feynman's ratchet in Refs.~\cite{ParrondoEspagnol96}). 
This is because momenta transfer heat in the timescale of momentum relaxation and therefore thermal conductivity between the reservoirs is inversely proportional to this timescale \cite{ParrondoEspagnol96}, which is assumed to be infinitesimal in the overdamped limit.
If alternatively we start from the Fokker-Planck approach where the fast momenta have been eliminated as in Refs.~\cite{EvB10_2}, we completely fail to evaluate the heat flows due to momentum transfer between the reservoirs. Indeed, contrary to the single reservoir case, the fast momenta are not in an equilibrium but in a nonequilibrium steady state and their contribution to the heat conduction must be accounted for.
Therefore, a more sophisticated method is needed to establish an overdamped description in presence of multiple reservoirs.

An important comment should be made at this point.
Many systems made of two or more interacting Brownian particles 
have been considered where each reservoir acts on a different particle (see, e.g., Refs.~\cite{Sekimoto, FogedbyImparato12, DMVO13,ChunNoh15}).
Various models of Feynman ratchets \cite{Feynman63} fall for instance into this category \cite{ParrondoEspagnol96, Sek97, MS98, GS05}.
In this case, the thermodynamic problems associated to the overdamped description do not occur because each momentum equilibrates with its own reservoir.

Our paper is organized as follows.
In Sec. II, we formulate stochastic thermodynamics for a system coupled to multiple reservoirs and described by an underdamped Langevin dynamics. 
We first establish the first and second law of thermodynamics. We then build a time-evolution equation for the heat generating function which we then use 
to derive the integral and the detailed fluctuation theorems.    
In Sec. III, we show that the naive overdamped descriptions fail to assess the heat flows.
Then, we derive the overdamped time-evolution equation for the heat generating function from the underdamped one. 
We do so by exploiting the timescale separation between momenta and positions which occurs in the limit of high friction.
We show that our overdamped theory is thermodynamically consistent and satisfies the fluctuation theorems.  
We also emphasize that it is not equivalent to the wrong prediction that one would obtain by naively starting from an overdamped Langevin or Fokker-Planck description.
In Sec. IV, we analytically solve the overdamped and underdamped heat statistics for a Brownian particle on a ring 
in contact with two reservoirs and subjected to a non-conservative force. We show that the long time cumulants for 
the heat transfer statistics of our overdamped theory are the same as those of the underdamped theory.   
Conclusions are drawn in Sec. V.
Various technical aspects of the paper are relegated to Appendices to improve its readability.

\section{Underdamped stochastic thermodynamics with multiple heat reservoirs}

In this Section, we introduce the underdamped stochastic thermodynamics in presence of multiple reservoirs, and investigate its properties.

\subsection{Underdamped Langevin dynamics and heat}

We start from the $N$-dimensional underdamped Langevin equation simultaneously coupled to multiple heat reservoirs:
\eqn{\label{ULeq1}
	d x_t &=& v_t dt,\\
	m d v_t &=& f_t(x_t)dt+\sum_\nu \left(-\gamma^\nu v_t dt + \sqrt{2\gamma^\nu\kb T^\nu}dw^\nu_t\right),\ \ \ \label{ULeq2}
}
where $f_t$ is the systematic force, which can be separated into the conservative force and the non-conservative force as
$
	f_t(x) = -\dd_x V_t(x) + f_t^{\rm nc}(x).
$
We define $\gamma^\nu$ and $T^\nu$ as the friction coefficient and the temperature of the $\nu$-th reservoir, respectively.
The Wiener processes $dw^\nu_t$ satisfy
$
	\ave{dw^\nu_t}= 0
$
and
$
	dw^\nu_t (dw^{\nu'}_t)^T = \delta^{\nu\nu'}\mathds{1} dt,
$
where $\mathds{1}$ is the $N\times N$ unit matrix.
Equations~(\ref{ULeq1}) and (\ref{ULeq2}) are dynamically equivalent to the effective equations
\eqn{\label{ULeq3}
	d x_t &=& v_t dt,\\
	m d v_t &=& f_t(x_t) dt -\gamma^{\rm ef} v_t dt + \sqrt{2\gamma^{\rm ef}\kb T^{\rm ef}} dw^{\rm ef}_t,
	\label{ULeq4}
}
where $dw^{\rm ef}_t$ is another Wiener process, and we define the effective friction coefficient as
$
	\gamma^{\rm ef} := \sum_\nu \gamma^\nu,
$
and the effective temperature as
$
	 T^{\rm ef} := {\sum_\nu \gamma^\nu T^\nu}/{\sum_\nu \gamma^\nu}
	 ={\sum_\nu \gamma^\nu T^\nu}/{\gamma^{\rm ef}}.
$
However, Eqs.~(\ref{ULeq1}) and (\ref{ULeq2}) are not thermodynamically equivalent to Eqs.~(\ref{ULeq3}) and (\ref{ULeq4}), because in Eq.~(\ref{ULeq4}) thermal noises from each reservoir are mixed and cannot be used anymore to discriminate the heat flow from each reservoir.

In accordance with the work by Sekimoto \cite{Sek97,Sekimoto}, we define heat flows from the system to the reservoirs by
\eqn{\label{SekH}
	dQ^\nu_t := -v_t\circ (-\gamma^\nu v_t dt + \sqrt{2\gamma^\nu\kb T^\nu}dw^\nu_t),
}
where the symbol $\circ$ means the Stratonovich product.
The Stratonovich product
$
	v_t \circ dw^\nu_t
$
can be transformed into the It\^o product as
$
	v_t \cdot dw^\nu_t+N\sqrt{2\gamma^\nu\kb T^\nu} dt/2m
$
(see, for example, Sec.~4.4 of Ref.~\cite{Gardiner}).
Therefore, in terms of the It\^o product, the heat flows are
\eqn{\label{TEofH}
	dQ^\nu_t = \frac{N\kb\gamma^\nu}{m} \left(
		\frac{m v_t^2}{N\kb}-T^\nu
	\right) dt
	-\sqrt{2\gamma^\nu \kb T^\nu} v_t\cdot dw^\nu_t.\nonumber\\
} 
When we take the average, we obtain
\eqn{
	\ave{dQ^\nu_t} = \frac{N\kb\gamma^\nu}{m} \left(
		\frac{m \ave{v_t^2}}{N\kb}-T^\nu
	\right) dt.
}
The terms in the parenthesis can be regarded as the difference between the effective temperature of the momentum degrees of freedom and the temperature of the $\nu$-th heat reservoir.

\subsection{First and second law}
The internal energy of the system can be defined as
$
	U_t(x,v) = V_t(x) + mv^2/2.
$
Then, its increment can be written as
\eqn{
	dU_t(x_t,v_t) = (\dd_t V_t (x_t)) dt + (\dd_x V_t(x_t)) dx_t + mv_t \circ dv_t.\nonumber\\
}
Using Eqs.~(\ref{ULeq1}-\ref{ULeq2}) and (\ref{SekH}), we obtain
\eqn{
	dU_t(x_t,v_t) =
	 dW^{\rm c}_t+ dW^{\rm nc}_t - \sum_\nu dQ^\nu_t.\ \ \ 
}
The first term on the right-hand side
$
	dW^{\rm c}_t:=(\dd_t V_t(x_t))dt
$
represents the conservative work, while the second term
$
	dW^{\rm nc}_t:=f^{\rm nc}_t(x_t)v_tdt
$
is the non-conservative work.
When we consider the evolution from time $t=0$ to $\tau$, we obtain
\eqn{\label{1stLaw}
	\Delta U = W^{\rm c}+W^{\rm nc}-\sum_\nu Q^\nu
	= W -\sum_\nu Q^\nu
}
where $\Delta U = U_\tau(x_\tau,v_\tau)-U_0(x_0,v_0)$, $I=\int_{t=0}^\tau dI_t$ for $I=W^{\rm c},W^{\rm nc},Q^\nu$, and $W=W^{\rm nc}+W^{\rm c}$.
In this way, we can obtain the first law of thermodynamics (\ref{1stLaw}) at the trajectory level.
Consequently, we also obtain the first law at the ensemble level as
\eqn{
	\ave{\Delta U} = \ave{W} - \sum_\nu \ave{Q^\nu}.
}

We define the stochastic Shannon entropy of the system by
$
	s_t := -\kb \ln P_t(x_t, v_t),
$
where $P_t$ is the probability distribution function at time $t$.
On the other hand, the entropy production of the $\nu$-th reservoir is 
$
	dQ^\nu_t/T^\nu.
$
Therefore, the total entropy production should be identified as
\eqn{\label{defstot}
	\Delta s^{\rm tot}
	:=
	\Delta s + \sum_\nu \frac{Q^\nu}{T^\nu},
}
where $\Delta s:=s_\tau-s_0$.
At the trajectory level, $\Delta s^{\rm tot}$ can be either positive or negative.
However, at the ensemble level, the second law of thermodynamics holds:
\eqn{\label{ud2law}
	\ave{\Delta s^{\rm tot}}\ge 0,
}
which can be derived from the integral fluctuation theorem presented in Section~II~D.

\subsection{Generating function and its time evolution}

We define the heat generating function as
\eqn{
	&&G_t(x,v,\{\Lambda^\nu\})\nonumber\\
	&&:=
	\ave{\delta(x_t-x)\delta(v_t-v)\exp\left[\sum_\nu \frac{\Lambda^\nu Q^\nu_t}{\kb T\ef} \right]}.
}
When all $\Lambda^\nu$ vanish, $G_t$ reduces to the probability distribution function
$
	G_t(x,v,\{0\}) =
	\ave{\delta(x_t-x)\delta(v_t-v)}
	=:P_t(x,v).
$
At the initial time $t=0$, all heats $Q^\nu_0$ vanish, and we therefore obtain
\eqn{
	G_0(x,v,\Lam) &=& \ave{\delta(x_0-x)\delta(v_0-v)} \nonumber\\
	&=& P_0(x,v),
}
which is the initial condition for the generating function.
Moreover, we define the integrated generating function by
\eqn{
	\mathcal G_t (\{\Lambda^\nu\})&=&
	\ave{\exp\left[\sum_\nu \frac{\Lambda^\nu Q^\nu_t}{\kb T\ef}\right]}\\
	&=& \int dx dv\ G_t(x,v,\{\Lambda^\nu\})
}
For convenience, we also define the cumulant generating function:
\eqn{
	\mC_t(\Lam) &:=& \ln \mG_t(\Lam).
}
From the cumulant generating function, we can obtain, for example, the average of the heats
$
	\ave{Q^\nu_t} =(\kb T\ef)
	\left.
		 {\dd \mC_t }/{\dd \Lambda^\nu}
	\right|_{\{\Lambda^\nu=0\}},
$
and the covariance
$
	\langle Q^\nu_t Q^{\nu'}_t\rangle
	-\langle Q^\nu_t\rangle \langle Q^{\nu'}_t \rangle =
	(\kb T\ef)^2
		 {\dd^2 \mathcal C_t }/{\dd \Lambda^\nu\dd \Lambda^{\nu'}}
	|_{\{\Lambda^\nu=0\}}.
$
In this way, we can obtain all the cumulants (moments) of heat by differentiating $\mC_t$ ($\mG_t$).
Thus, $\mC_t$ or $\mathcal G_t$ have the equivalent amount of information to the joint probability distribution of heat $P(\{Q^\nu_t\})$.

Using Eqs.~(\ref{ULeq1}), (\ref{ULeq2}) and (\ref{TEofH}), we obtain the time evolution of the generating function as
\eqn{\label{udteeq}
	\dd_t G_t = \mL_t(\Lam)G_t,
}
where
\eqn{\label{teop}
	\mL_t(\Lam) = \mL^0_t + \frac{1}{\epsilon}\mL^1(\Lam),
}
and $\epsilon:=m/\gamma^{\rm ef}$ is the overdamped parameter, which characterizes the timescale of momentum relaxation and we assume to be small.
Here, we define
\eqn{
	\mL^0_t
	&=& - \tilde v\dd_{\tilde x} - \tilde f_t(\tilde x)\dd_{\tilde v},\\
	\mL^1(\Lam)
	&=&
	\dd_{\tilde v}^2 + (1+2 B) \tilde v\dd_{\tilde v} + (A\tilde v^2+NB+N).\nonumber\\
}
The variables with tildes are rescaled as detailed in Appendix A.
The derivation of the time-evolution equation is given in Appendix B.
The coefficients $A$ and $B$ in $\mL^1$ are functions of counting fields $\Lam$ defined by
\eqn{
	A(\{\Lambda^\nu\})&=&\sum_\nu \Lambda^\nu {\tilde \gamma}^\nu (1+\Lambda^\nu {\tilde T}^\nu),\\
	B(\{\Lambda^\nu\})&=&\sum_\nu \Lambda^\nu {\tilde \gamma}^\nu {\tilde T}^\nu,
}
where ${\tilde \gamma}^\nu=\gamma^\nu/\gamma^{\rm ef}$ and ${\tilde T}^\nu=T^\nu/T^{\rm ef}$ are the relative friction coefficient and the relative temperature, respectively.
Therefore, we note that when we set all the counting fields to zero ($\{\Lambda^\nu=0\}$), Eq.~(\ref{udteeq}) reduces to the Kramers equation
\eqn{\label{Keq}
	\dd_t P_t
	=
	-\tilde v\dd_{\tilde x}P_t
	-\tilde f_t\dd_{\tilde v}P_t
	+\frac{1}{\epsilon}[\dd_{\tilde v}^2 P_t
	+\dd_{\tilde v}(\tilde v P_t)].
}

\subsection{Fluctuation theorems}
Here, we enumerate fluctuation theorems in underdamped stochastic thermodynamics.
The derivations are relegated to Appendix C.

For a finite time interval and any initial condition that vanished nowhere in the phase space, the integral fluctuation theorem for the total entropy production~(\ref{defstot}) holds
\eqn{\label{udift}
	\langle e^{-\Delta s^{\rm tot}/\kb} \rangle = 1.
}
Consequently, thanks to the Jensen inequality, we obtain the second law of thermodynamcis
\eqn{
	\langle \Delta s^{\rm tot} \rangle \ge 0.
}
For an initial condition that vanishes in a certain region, we can modify the integral fluctuation theorem~(\ref{udift}) and confirm that the second law is still valid \cite{MFU14}.

Next, we consider the situation where the system starts from a thermal equilibrium state $P^{\rm eq}_0$ corresponding to the reference reservoir $\nu=0$,
where the instantaneously equilibrium state at time $t$ with the temperature $T^0$ is defined as
\eqn{
	P^{\rm eq}_t(x,v)
	=
	\exp\left[
		-\frac{U_t(x,v)-F_t}{\kb T^0}
	\right].
}
The equilibrium free energy $F_t$ is given by
\eqn{
	F_t = -\kb T^0
	\ln \int dx dv\ 
	\exp\left[
		-\frac{U_t(x,v)}{\kb T^0}
	\right].
}
Then, in accordance with Ref.~\cite{CEI14}, the irreversible entropy production should be identified as
\eqn{\label{irrEP}
	\Delta_{\rm i} s
	&=&
	-\kb\ln P^{\rm eq}_\tau(x_\tau,v_\tau)
	+\kb\ln P^{\rm eq}_0(x_0,v_0)
	+\sum_\nu \frac{Q^\nu}{T^\nu}\nonumber\\
	&=&
	\frac{W-\Delta F -\sum_\nu \eta^\nu Q^\nu}{T^0},
}
where 
$
	\eta^\nu=1-T^0/T^\nu
$
is the Carnot efficiency between the reference reservoir and the $\nu$-th reservoir.
For $\Delta_{\rm i} s$, we can derive the detailed fluctuation theorem
\eqn{\label{uddft}
	\frac{\bar P(-\Delta_{\rm i} s)}{P(\Delta_{\rm i} s)}
	=
	\exp [-\Delta_{\rm i} s/\kb],
}
where $\bar P$ represents the probability in the reversed process where the initial state is the equilibrium state $P^{\rm eq}_\tau$ and the explicit time dependence of the forces is time-reversed.
Hence, we obtain the second law
\eqn{
	\langle \Delta_{\rm i} s \rangle \ge 0.
}

Finally, we mention that the generator $\mL_t$ can be shown to have the fluctuation-theorem symmetry
\eqn{\label{udssft}
	\mL^\dag_{t,v\to -v}(\{-\Lambda^\nu-\beta^\nu\})
	=
	\mL_t(\Lam).
}
Unfortunately, this does not imply a steady-state fluctuation theorem for the joint probability 
distribution of heats because of the non-analyticity of the generating function.
The details are provided in Appendix C.

We stress that it is the first time that the above-mentioned fluctuation theorems are explicitly derived for underdamped Langevin dynamics in presence of multiple reservoirs.

\section{Overdamped approximation}
In this Section, we show how two naive overdamped descriptions fail to properly evaluate the heat flows in presence of multiple reservoirs.
Then, starting from the underdamped theory described in the previous section, we derive the correct overdamped approximation of Eq.~(\ref{udteeq}) by utilizing timescale separation techniques and singular expansion.

\subsection{Naive overdamped descriptions fail}
First, we try to start from the overdamped Langevin equation coupled to two reservoirs with temperatures $T^{\rm h}$ and $T^{\rm c}$ ($T^{\rm h}>T^{\rm c}$).
For simplicity, we assume vanishing systematic force:
\eqn{\label{naiveLeq}
	0=-\gamma^{\rm h} \dot x_t -\gamma^{\rm c} \dot x_t + \zeta^{\rm h}_t + \zeta^{\rm c}_t.
}
The white Gaussian noises $\zeta^\nu\ (\nu={\rm h},{\rm c})$ satisfy
$
	\langle \zeta^\nu_t \rangle= 0,
	\langle \zeta^\nu_t\zeta^{\nu'}_s \rangle = 2\delta^{\nu\nu'}\gamma^\nu \kb T^\nu \delta (t-s).
$
Naively applying Sekimoto's heat definition \cite{Sek97, Sekimoto}, the heats flowing from the system to the reservoirs from time $t=0$ to $\tau$ should be defined as
\eqn{\label{naiveSekH}
	Q^\nu = -\int_0^\tau dt\ \dot x_t (-\gamma^\nu \dot x_t + \zeta^\nu_t).
}
Using Eq.~(\ref{naiveLeq}), we can formally obtain the averaged heats as
\eqn{
	\langle Q^{\rm h} \rangle =
	-\frac{2\kb\gamma^{\rm h}\gamma^{\rm c}}{(\gamma^{\rm h}+\gamma^{\rm c})^2}(T^{\rm h}-T^{\rm c}) \int_0^\tau dt\ \delta (0)
	=
	-\langle Q^{\rm c} \rangle.\ \ \ 
}
Therefore, $\langle Q^{\rm h} \rangle$ ($\langle Q^{\rm c}\rangle$) is negatively (positively) divergent and therefore ill-defined.

As a second attempt, we try to construct stochastic thermodynamics from the overdamped Fokker-Planck equation with additive currents.
According to Ref.~\cite{EvB10_2}, the Fokker-Planck equation for the overdamped probability distribution function $P\od(x)$ in presence of the multiple reservoirs reads
\eqn{\label{naiveFP}
	\dot P\od_t(x)
	=
	-\dd_x J\od_t(x),
}
where $J\od_t(x)$ is the sum of the current due to each reservoir as
$
	J\od_t(x)
	=
	\sum_\nu J^{\rm od,\nu}_t(x)
$
and
\eqn{
	J^{\rm od,\nu}_t(x)
	=
	\frac{1}{\gamma^\nu} (f_t(x)-\kb T^\nu \dd_x) P\od_t(x).
}
We note that the Fokker-Planck equation~(\ref{naiveFP}) can be obtained as the continuous limit of the master equation in presence of multiple reservoirs \cite{EvB10}.
In this scheme, as elaborated in Ref.~\cite{EvB10_2}, the averaged heat flowing from the system to the $\nu$-th reservoir should be identified as
\eqn{
	\langle \dot Q^\nu_t \rangle
	=
	\int dx\ J^{\rm od,\nu}_t(x) f_t(x).
}
As a result, we observe that there are no heat flows on average in the absence of systematic force, namely when $f_t(x)=0$.
Once again, this result is unphysical because momenta are not in an equilibrium state but in a nonequilibrium steady state and should therefore transfer heat between the reservoirs with different temperatures.

In the following Sections, we utilize the singular expansion of the time-evolution equation of the underdamped heat generating function and derive an alternative overdamped description, which correctly evaluates the  heat flows in presence of multiple reservoirs.

\subsection{Timescale separation leads to overdamped description}
Here, we eliminate the fast degrees of freedom, i.e., momenta, and obtain a time-evolution equation of positions.
To this aim, we introduce fast and slow timescales and conduct singular expansion of Eq.~(\ref{udteeq}) with respect to $\epsilon$.
A similar method is used in Ref.~\cite{BC14}.
The details of derivation are given in Appendix D.

We define fast timescale $\theta=\epsilon^{-1}t$, intermediate timescale $t$ and slow timescale ${\hat t}=\epsilon t$ and deal them as independent variables.
We assume that the time variation of the force is in the intermediate and the slow timescale as $f=f_{t,{\hat t}}$.
In other words, time-dependent driving is assumed to be slower than the momentum relaxation timescale $\theta$.
Then, the time-evolution equation~(\ref{udteeq}) reads
\eqn{\label{tsste}
	&&\frac{1}{\epsilon}\dd_\theta G_{\theta,t,{\hat t}}
	+ \dd_t G_{\theta,t,{\hat t}}
	+ \epsilon \dd_{{\hat t}} G_{\theta,t,{\hat t}}\nonumber\\
	&&=
	\mL^0_{t,{\hat t}} G_{\theta,t,{\hat t}}
	+ \frac{1}{\epsilon} \mL^1(\Lam) G_{\theta,t,{\hat t}}.
}
We assume that we can expand $G$ with respect to $\epsilon$ as
\eqn{
	G_{\theta,t,{\hat t}}
	=
	G^{(0)}_{\theta,t,{\hat t}}
	+\epsilon G^{(1)}_{\theta,t,{\hat t}}
	+ \epsilon^2 G^{(2)}_{\theta,t,{\hat t}}+\cdots
}
Then, from each order of $\epsilon$, we obtain
\eqn{
	(\dd_\theta-\mL^1) G^{(0)}_{\theta,t,{\hat t}}
	&=&
	0,\label{eq-1}\\
	(\dd_\theta-\mL^1) G^{(1)}_{\theta,t,\hat t}
	&=&
	-(\dd_t-\mL^0_{t,\hat t}) G^{(0)}_{\theta,t,\hat t},
	\label{eq0}\\
	(\dd_\theta-\mL^1) G^{(2)}_{\theta,t,\hat t}
	&=&
	-(\dd_t-\mL^0_{t,\hat t}) G^{(1)}_{\theta,t,\hat t}
	-\dd_{\hat t} G^{(2)}_{\theta,t,\hat t},
	\label{eq1}
}
and higher order equalities.
Then, we solve these equations order by order.
As explained in Appendix D, $\dd_\theta$ can be replaced by the largest eigenvalue $\alpha_0(\Lam)$ of $\mL^1(\Lam)$ after relaxation in the fast timescale.
Consequently, from Eq.~(\ref{eq-1}), we conclude that
\eqn{
	G^{(0)}_{\theta,t,\hat t}(x,v,\Lam)
	=
	\hat G^{(0)}_{t,\hat t}(x,\Lam)e^{\alpha_0(\Lam) \theta} \phi_0(v,\Lam),\nonumber\\
}
where $\phi_0(v,\Lam)$ is the right eigenfunction of $\mL^1(\Lam)$ corresponding to $\alpha_0(\Lam)$ and $\hat G^{(0)}_{t,\hat t}(x,\Lam)$ is an arbitrary function independent from $\theta$ and $v$.
We can observe that the left-hand side of Eq.~(\ref{eq0}) is orthogonal to the left eigenfunction $\bar\phi_0(v,\Lam)$ of $\alpha_0(\Lam)$.
As a result, we obtain the following equation for $\hat G^{(0)}$ as the solvability condition for Eq.~(\ref{eq0}):
$
	\dd_t \hat G^{(0)}_{t,\hat t}(x,\Lam) = 0.
$
Under this condition, Eq.~(\ref{eq0}) can be explicitly solved and $G^{(1)}$ can be written in terms of another arbitrary function $\hat G^{(1)}$,
which is in turn confirmed to satisfy the solvability condition for Eq.~(\ref{eq1}).
The original generating function can be therefore approximated as
\eqn{
	\mG_t(\Lam)
	&=&
	\mG_t^{v}(\Lam)
	\int dx\ [G^{\rm od}_t(x,\Lam) + \mO(\epsilon^2)],\ \ \ \ \ 
}
where the contribution of the momentum degrees of freedom is represented by
\eqn{\label{odGv}
	\mG_t^{v}(\Lam)
	=
	\exp\left[\frac{\alpha_0(\Lam) t}{\epsilon}\right]
}
and the overdamped generating function is given by
\eqn{
	G^{\rm od}_t(x,\Lam)
	\propto
	\hat G^{(0)}_{\hat t}(x,\Lam)
	+\epsilon \hat G^{(1)}_t(x,\Lam).
}
Hence, we can regard $G\od$ as the overdamped part of the generating function since $G\od$ is independent from $v$.

\subsection{Overdamped approximation}
As elaborated in Appendix D, using the solvability condition for Eq.~(\ref{eq1}),
one can show that the time evolution of $G\od$ is given by
\eqn{\label{odteeq}
	\dd_t G\od_t
	=
	\mL\od_t(\Lam)G\od_t
	+ \mO(\epsilon^2),
}
where the overdamped time-evolution operator is
\begin{widetext}
\eqn{\label{odteop}
	\mL\od_t(\Lam)
	=
	\frac{1}{\gamma\ef R^2}
	\left[
		\kb T\ef\dd_x^2
		- \kappa\dd_x(f_t(x)\cdot)
		+ \rho f_t(x)\dd_x
		+ \frac{A}{\kb T\ef}(f_t(x))^2
	\right],
}
\end{widetext}
where $R=\sqrt{(1+2B)^2-4A}$, $\kappa=(1+2B+R)/2$ and $\rho=(-1-2B+R)/2$ are functions of the counting fields $\Lam$.
The initial condition is given by
$
	G\od_0(x,\Lam)
	=
	P\od_0(x),
$
where $P\od$ is the marginal probability distribution of the positions:
$
	P\od_t(x) = \int dv P_t(x,v).
$
We define
$
	\mG\od_t(\Lam) = \int dx\ G_t\od(x,\Lam).
$
Then, the underdamped generating function can be written as
\eqn{\label{odapprox}
	\mG_t(\Lam)
	=
	\mG_t^{v}(\Lam)
	[\mG\od_t(\Lam) + \mO(\epsilon^2)],
}
where
$
	\mG_t^{v}(\Lam)
	=
	\exp[N(1-R)t/(2\epsilon)]
$
because $\alpha_0=N(1-R)/2$.
Using the time evolution~(\ref{odteop}), we can completely calculate the overdamped generating function $G\od_t$ and
the approximate moments of heat can be obtained by differentiating Eq.~(\ref{odapprox}).
This is our main result.
We note that when we set $\{\Lambda^\nu=0\}$ the time-evolution operator of generating function reduces to the Fokker-Planck operator:
\eqn{
	\mL\od_t(\{\Lambda^\nu=0\})
	=
	-\frac{1}{\gamma\ef}\dd_x(f_t(x)\cdot)
	+ \frac{k_{\rm B}T\ef}{\gamma\ef} \dd_x^2.
}
As a consistency check, when we set $\tilde T^\nu=1,\ \Lambda^\nu=\Lambda$ for all $\nu$, we find that $\alpha_0=0$ and $\mL\od_t$ reduces to time-evolution operator of the heat generating function derived from the Langevin equation with a single reservoir (see Appendix E).

Let us furthermore emphasize that going through our systematic procedure is indispensable.
Equation~(\ref{odapprox}) is not simply the product of the generating function for the fast momentum degrees of freedom~(\ref{odGv}) and the heat generating function that one would naively derive by extending the overdamped heat generating function from one to multiple reservoirs.
The dynamics of the latter, contrary to Eq.~(\ref{odteop}), would be made of additive contributions from each reservoir.
This shows that the heat transfers due to momenta nontrivially affect those due to positions.

We now consider the error of our approximation.
By differentiating Eq.~(\ref{odapprox}) $n$ times, we can obtain the $n$-th moment of the heats.
The largest error comes from the term
\eqn{
	\mathcal{O}(\epsilon^2) \frac{\dd^n}{\dd \Lambda^{\nu^1}\cdots\dd \Lambda^{\nu^n}}\exp\left[
		\frac{\alpha_0(\Lam)}{\epsilon}t
	\right],
}
which is $\mathcal{O} (\epsilon^{2-n})$.
Thus, the $n$-th moment obtained from our approximation contains an error of $\mathcal{O} (\epsilon^{2-n})$.
The cumulant generating function $\mC_t$ can be approximated as
\eqn{\label{Capprox}
	\mC_t \simeq \frac{\alpha_0}{\epsilon}t+\ln \mG\od_t,
}
where the symbol $\simeq$ represents the approximate equality.
The first term represents heat conduction from the fast degrees of freedom on the fast timescale of $\theta$, while the second term represents the heat generated on the slow timescale ${\hat t}$, because the evolution of Eq.~(\ref{odteop}) is on this timescale.

Next, we mention how to implement boundary conditions in our overdamped approximation.
In the underdamped theory, the reflecting boundary condition at the position $x_0$ is written as
\eqn{\label{RBC}
	G_t(x_0,v,\Lam)=G_t(x_0,\mR v,\Lam),
}
where
$
	\mR v = v - 2(v\cdot \hat n) \hat n
$
is the velocity reflected by the boundary, whose outward-facing normal vector is $\hat n$.
The zeroth-order term~(\ref{G0}) obviously satisfies the symmetry~(\ref{RBC}).
The first-order term~(\ref{G1}) has the symmetry~(\ref{RBC}), when 
$
	\hat n(\dd_x-\kappa f_t) \tilde G^0_{{\hat t}}(x,\Lam)=0.
$
Therefore, the reflecting boundary condition is approximated as
\eqn{
	\hat n(\dd_x-\kappa f_t) G^{\rm od}_t(x,\Lam)= 0.
}

On the other hand, the absorbing boundary condition $G_t(x_0,v,\Lam)=0$ in the underdamped theory is translated as 
$
	G^{\rm od}_t(x_0,\Lam)=0
$
in the overdamped theory.

We now comment on why our method circumvents the heat divergence problem encountered in the naive Langevin approach in Sec.~III~A.
We have to take the limit $\epsilon\to+0$ to obtain Eq.~(\ref{naiveLeq}), which means that the relaxation time of momenta is infinitesimal.
In presence of reservoirs with multiple temperatures, momenta are in a nonequilibrium steady state and conduct heat between different reservoirs.
Since the relaxation time is infinitesimal, these heat flows are instantaneously transported.
This is the reason why they diverge when we naively extend Sekimoto's overdamped heat definition as in Eq.~(\ref{naiveSekH}) to multiple reservoirs.
In contrast, the heat flows evaluated using our method remain finite because we keep $\epsilon$ small but nonzero.

\subsection{First and second law}
We here consider the first and second law in our approximation.
Since $G\od$ is the generating function, we can introduce its corresponding stochastic heats $Q^{\rm od,\nu}$ satisfying
\eqn{
	G\od_t(x,\Lam)
	=
	\left\langle
		\delta(x_t-x)
		\exp\left[
			\sum_\nu \frac{\Lambda^\nu Q^{\rm od,\nu}_t}{\kb T\ef}
		\right]
	\right\rangle.
}
We also define the stochastic heats $Q^{v,\nu}$ transferred by the momenta by
\eqn{
	\mG^v_t(\Lam)
	=
	\left\langle
		\exp\left[
			\sum_\nu \frac{\Lambda^\nu Q^{v,\nu}}{\kb T\ef}
		\right]
	\right\rangle.
}

If we set all $\Lambda^\nu$ to be equal, since $\alpha_0(\{\Lambda^\nu=\Lambda\})=0$, we obtain
\eqn{
	\mG^v_t(\{\Lambda^\nu=\Lambda\})
	=
	\left\langle
		\exp\left[
			\frac{\Lambda}{\kb T\ef} \sum_\nu Q^{v,\nu}_t
		\right]
	\right\rangle
	=
	1,
}
which indicates that the sum of heats transferred by the momenta is always vanishing:
\eqn{
	\sum_\nu Q^{v,\nu}_t = 0.
}
Therefore, the heats transferred in the fast timescale completely balance each other without any interference from work or heat flows in slower timescales.
In particular, we note that the non-conservative work $dW^{\rm nc}_t=f^{\rm nc}_t(x_t)v_tdt$ is generated in the slow timescale because $v_t\simeq f_t(x_t)/\gamma\ef=\mathcal{O}(\epsilon)$ after the fast momentum relaxation.
On the other hand, from Eq.~(\ref{odapprox}), we obtain
\eqn{
	\mG_t(\{\Lambda^\nu=\Lambda\})
	\simeq
	\mG\od_t(\{\Lambda^\nu=\Lambda\}),
}
which implies that
\eqn{
	\sum_\nu Q^\nu_t
	\simeq
	\sum_\nu Q^{\rm od, \nu}_t.
}
This means that the total heat current is generated only by the overdamped heats.
Therefore, the first law~(\ref{1stLaw}) reduces to
\eqn{
	\Delta U
	\simeq
	W-\sum_\nu Q^{\rm od, \nu}_t.
}

We now turn to the second law.
We can define the overdamped total entropy production as
\eqn{
	\Delta s^{\rm od,tot}
	=
	\Delta s\od + \sum\frac{Q^{\rm od,\nu}}{T^\nu},
}
where $\Delta s\od$ is the difference of the overdamped Shannon entropy $s\od_t = -\kb \ln P\od_t(x_t)$.
By using the fluctuation theorem presented in the next Section, one can show that this quantity satisfies the inequality
\eqn{
	\langle \Delta s^{\rm od,tot} \rangle \ge 0.
}

\subsection{Fluctuation theorems}
To demonstrate the consistency of our approach, we show that the fluctuation theorems in the original underdamped theory are still valid in our overdamped theory.
The detailed derivations are provided in Appendix F.


For any initial condition that vanishes nowhere, one can derive the integral fluctuation theorem
\eqn{\label{odift}
	\langle e^{-\Delta s^{\rm od,tot}/\kb}\rangle = 1,
}
which is the overdamped analog of the integral fluctuation theorem~(\ref{udift}).

Next, we consider a situation where the system starts from equilibrium with respect to the reference reservoir $P_0^{\rm od,eq}(x)$, where
\eqn{
	P^{\rm od,eq}_t (x) = \exp\left[
		-\frac{V_t(x)-F\od_t}{\kb T^0}
	\right]
}
and
$
	F\od_t
	=
	-\kb T^0 \ln
	\int dx\ \exp\left[
		- {V_t(x)}/{\kb T^0}
	\right].
$
The overdamped version of the irreversible entropy production can defined by
\eqn{\label{EPoverFT}
	\Delta_{\rm i} s^{\rm od}
	=
	\frac{
		W-\Delta F\od - \sum_\nu \eta^\nu Q^{\rm od, \nu}
	}{T^0}.
}
However, an explicit expression for the non-conservative work $dW^{\rm nc}=f^{\rm nc}_t(x_t)v_tdt$ 
is out of reach since velocities $v_t$ have been eliminated in the overdamped approximation.
Therefore, to calculate the statistics of (\ref{EPoverFT}), we need to extend the overdamped 
approximation which we used for the heats statistics to also account for the non-conservative 
work statistics, as detailed in Appendix F.
By doing so, it can be shown that $\Delta_{\rm i} s\od$ satisfies the detailed fluctuation theorem
\eqn{\label{oddft}
	\frac{\bar P(-\Delta_{\rm i}s\od)}{P(\Delta_{\rm i}s\od)}
	=
	\exp[-\Delta_{\rm i} s\od /\kb],
}
which corresponds to Eq.~(\ref{uddft}).

Finally, we can also derive the fluctuation-theorem symmetry
\eqn{\label{odssft}
	\mL\od(\{-\Lambda^\nu-\beta^\nu\})
	=
	\mL\od(\Lam)
}
which is the twin of Eq.~(\ref{udssft}).

\subsection{Sufficient condition for equivalence}
We now show that the absence of a non-conservative force, $f^{\rm nc}=0$, is a sufficient condition 
under which the overdamped approximation asymptotically coincides with the underdamped results.

No non-conservative force means that the sole applied force is time-independent and conservative and can thus be written as
$
	f(x)=-\dd_x V(x).
$
In this case, the largest eigenvalue of the underdamped operator~(\ref{teop}) is
$
	{N(1-R)}/({2\epsilon})=\alpha_0/\epsilon,
$
and its corresponding eigenfunction is
\eqn{
	\exp\left[
		-\kappa\left(
			\frac{1}{2}v^2 + V(x)
		\right)
	\right].
}
On the other hand, the largest eigenvalue of the overdamped operator~(\ref{odteop}) is
$
	0,
$
and the corresponding eigenfunction is
$
	\exp\left[
		-\kappa V(x)
	\right].
$
Therefore, from Eq.~(\ref{GGod}), we can conclude that the dominant eigenvalue of $\mG_t$ calculated by the overdamped approximation is
$
	\alpha_0/\epsilon.
$
This thus proves that our overdamped approximation and the exact underdamped theory reproduce the same results asymptotically.
In this case, all the heat flows are asymptotically due to the momentum degrees of freedom.

In the next Section, we demonstrate the same asymptotic equivalence for a specific model in presence of constant non-conservative force.

\section{Analytically solvable model}
\begin{figure}
	\includegraphics{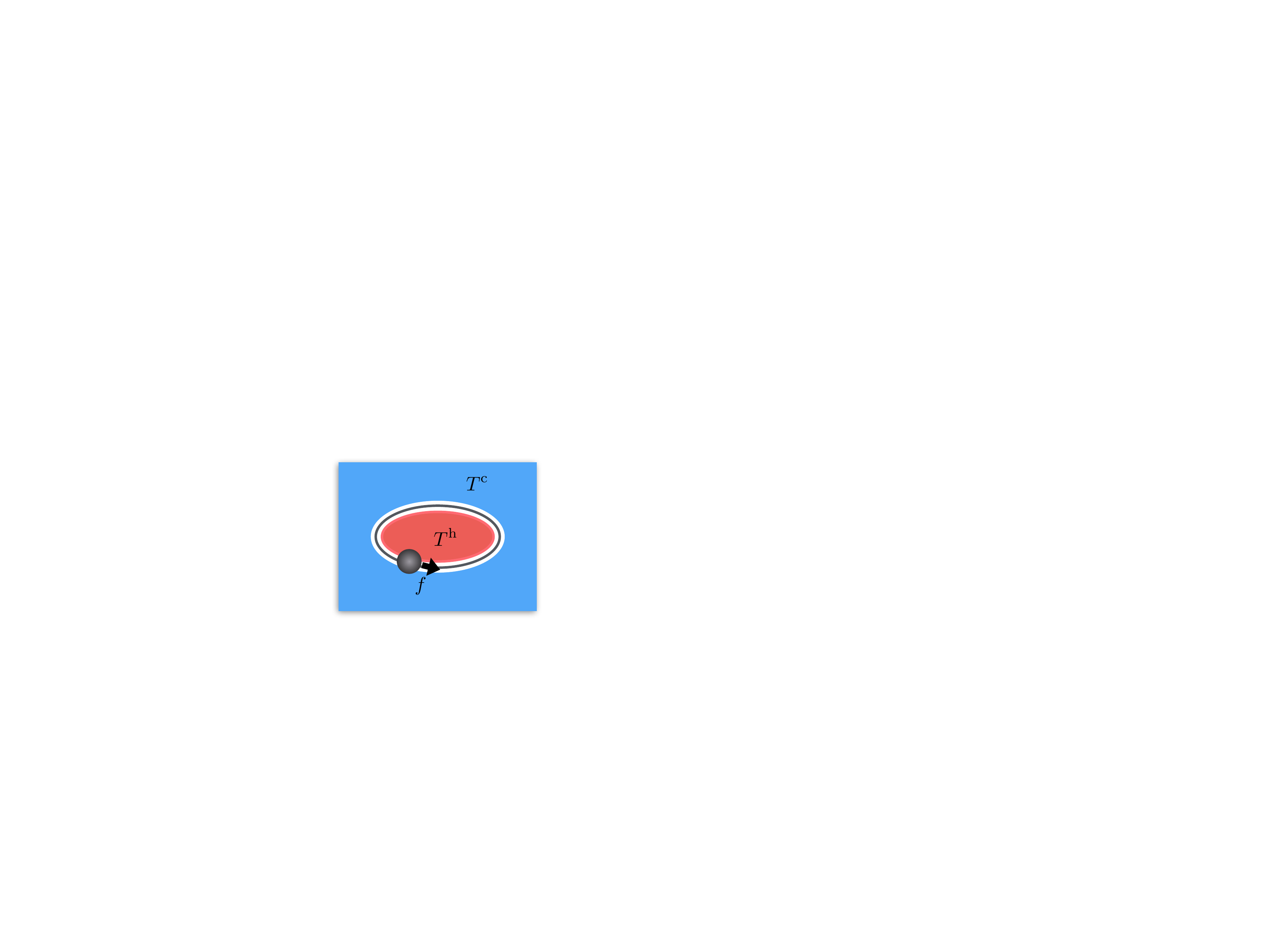}
	\caption{\label{RingWithTwoReservoirs}
		A Brownian particle is confined in a one-dimensional ring coupled to two heat reservoirs with temperatures $T^{\rm h}$ and $T^{\rm c}$. A constant external force $f$ is applied to the particle.
	}
\end{figure}

We consider a Brownian particle in contact with two heat reservoirs at temperatures 
$T^{\rm h}$ and $T^{\rm c}$ and confined on a one-dimensional ring of unit length. 
The particle is driven by a constant force $f$, as illustrated in Fig.~\ref{RingWithTwoReservoirs}. 
The system is assumed to be at steady state from the beginning.
This model is an analytically solvable model of a stochastic heat engine.

\subsection{Overdamped approximation}
From the rotational symmetry of the system and of its initial condition (here the steady state solution), we can 
ignore the position dependence of the generating function and our overdamped time evolution~(\ref{odteeq}) reduces to
\eqn{
	\dd_t G\od_t (\Lam)
	\simeq
	\frac{A f^2}{\gamma\ef\kb T\ef R^2} G\od_t,
}
with the initial condition
$
	G\od_0 (\Lam) = 1.
$
Therefore, the overdamped generating function is
\eqn{
	\mG\od_t
	\simeq
	\exp\left[
		\frac{A f^2}{\gamma\ef\kb T\ef R^2} t
	\right].
}
From Eq.~(\ref{Capprox}), the cumulant generating function is calculated as
\eqn{\label{exodcgf}
	\mC_t
	\simeq
	\frac{1-R}{2\epsilon}t + \frac{A f^2}{\kb\gamma\ef T\ef R^2} t.
}
In this way, we can easily obtain the generating functions in our overdamped approximation.
By differentiating this cumulant generating function, we obtain
\eqn{\label{exodave}
	\ave{Q^\nu_t}
	\simeq
	\frac{\kb \gamma^\nu}{m}(T\ef-T^\nu)t
	+ \frac{\gamma^\nu f^2}{(\gamma\ef)^2} t
}
The first term is the heat conduction due to the momentum degrees of freedom, which is proportional 
to the difference between the temperature of the $\nu$-th reservoir and the effective temperature.
The second term is the additional heat originating from the applied force.
The variance in turn is given by
\eqn{\label{exodvar}
	\frac{\ave{(Q^\nu_t)^2}-(\ave{Q^\nu_t})^2}{(\kb T\ef)^2}
	&\simeq&
	\frac{2}{\epsilon}{\tilde \gamma}^\nu({\tilde \gamma}^\nu+{\tilde T}^\nu-2{\tilde \gamma}^\nu{\tilde T}^\nu)t\nonumber\\
	&&\ \ \ +2\epsilon {\tilde \gamma}^\nu \tilde f^2 (4{\tilde \gamma}^\nu + {\tilde T}^\nu -4 {\tilde \gamma}^\nu{\tilde T}^\nu)t.\nonumber\\
}

\subsection{Direct calculation in the underdamped theory}
In the underdamped regime, we directly calculate the heat generating function.
The underdamped time-evolution equation~(\ref{udteeq}), reduces to
\eqn{\label{exteeq}
	\dd_t G_t(v,\Lam)=\mL_v (\Lam) G_t(v,\Lam),
}
where we define
\eqn{\label{exteop}
	\mL_v &=& -\frac{f}{m}\dd_{v}\nonumber\\
	&&+ \frac{1}{\epsilon}
	\left[
		\frac{\kb T\ef}{m}\dd_{v}^2
		+ (1+2B)\dd_{v}(v\cdot)
		+\frac{m}{\kb T\ef}A{v}^2
		-B
	\right].\nonumber\\
}
The steady-state distribution
\eqn{
	P^{\rm ss}(v)=\sqrt\frac{m}{2\pi\kb T\ef}
	\exp\left[
		-\frac{m}{2\kb T\ef} 
		\left(v-\frac{f}{\gamma\ef}\right)^2
	\right] 
}
is taken as the initial condition
\eqn{
	G_0(v,\Lam) = P^{\rm ss}(v).
}
As described in Appendix G, $\mL_v$ is essentially Hermite's differential operator.
Therefore, we can explicitly calculate the eigenvalues and the eigenfunctions of $\mL_v$.
Because we can expand $G_0=P^{\rm ss}$ in terms of Hermite's polynomials,
we can obtain explicit time evolution of the generating function $G_t$,
which can be tidied up by some formulae of Hermite's polynomials.

As a result, by differentiating the explicit expression of $\mG_t(\Lam)=\int dv\ G_t(v,\Lam)$, 
we can obtain the average of the heat flows
\eqn{
	\ave{Q^\nu_t}
	&=&
	\frac{\kb\gamma^\nu}{m}(T\ef-T^\nu)t
	+ \frac{\gamma^\nu f^2}{(\gamma\ef)^2} t,
}
which exactly coincides with the overdamped result (\ref{exodave}).
In turn the variance is given by
\eqn{
	\frac{\ave{(Q^\nu_t)^2}-\ave{Q^\nu_t}^2}{(\kb T\ef)^2}
	&=&
	\frac{2}{\epsilon }{\tilde \gamma}^\nu({\tilde \gamma}^\nu+{\tilde T}^\nu-2 {\tilde \gamma}^\nu {\tilde T}^\nu  )t
	\nonumber\\
	&&\ \ \ \ \ 
	+({\tilde \gamma}^\nu)^2(e^{-2t/\epsilon}-1)(1-2{\tilde T}^\nu)
	\nonumber\\
	&&\ \ \ \ \ 
	+2\epsilon {\tilde \gamma}^\nu \tilde f^2  ( 4 {\tilde \gamma}^\nu+ {\tilde T}^\nu -4 {\tilde \gamma}^\nu {\tilde T}^\nu)t
	\nonumber\\
	&&\ \ \ \ \ 
	+8\epsilon^2 ({\tilde \gamma}^\nu)^2 \tilde f^2 (e^{-2t/\epsilon} -1)(1-{\tilde T}^\nu).\nonumber\\
}
We observe that the overdamped approximation~(\ref{exodvar}) has an error of $\mathcal{O}(\epsilon^0)$ 
as expected from the discussion in Section~III~C.

Finally, we consider the asymptotic limit.
From the expression of $\mC_t$ not explicitely shown here, we can obtain the asymptotic cumulant generating function $\Phi_0$ as
\eqn{\label{exacgf}
	\Phi_0
	=
	\lim_{t\to\infty}\frac{1}{t}\mC_t
	=
	\frac{1-R}{2\epsilon} + \frac{A f^2}{\gamma\ef\kb T\ef R^2},
}
which coincides with the overdamped result~(\ref{exodcgf}).
As a result, we obtain
\eqn{
	\lim_{t\to\infty} \frac{\ave{Q^\nu_t}}{t}
	&=&
	\frac{\kb\gamma^\nu}{m}(T\ef-T^\nu)
	+ \frac{\gamma^\nu f^2}{(\gamma\ef)^2}\\
	\lim_{t\to\infty} \frac{1}{t}
	\frac{\ave{(Q^\nu_t)^2}-\ave{Q^\nu_t}^2}{(\kb T\ef)^2}
	&=&
	\frac{2}{\epsilon}{\tilde \gamma}^\nu({\tilde \gamma}^\nu+{\tilde T}^\nu-2{\tilde \gamma}^\nu{\tilde T}^\nu)\nonumber\\
	&&\ \ \ +2\epsilon {\tilde \gamma}^\nu \tilde f^2 (4{\tilde \gamma}^\nu + {\tilde T}^\nu -4 {\tilde \gamma}^\nu{\tilde T}^\nu),\nonumber\\
}
which coincide with the overdamped results (\ref{exodave}) and (\ref{exodvar}).

\section{Summary and Conclusions}

We begin by summarizing our findings.
We first established the stochastic thermodynamics of underdamped Brownian particles in contact with multiple reservoirs.
We derived the equation of motion for the generating function ruling the statistics of the heat transfers and used 
it to derive an integral and a finite-time fluctuation theorem.

We then showed that the overdamped Langevin equation or the Fokker-Planck equation cannot be used as a starting 
point to establish an overdamped theory of stochastic thermodynamics in presence of multiple reservoirs.
The reason is that the fast momentum variables which have been eliminated from the description play a crucial role in the heat transfers. 
Hence, we proposed a correct overdamped stochastic thermodynamics that we systematically derived by applying 
timescale separation techniques to the equation of motion for the generating function of the heat transfers.
We showed that it preserves the fluctuation theorems and that its prediction for the long-time 
heat statistics always coincides with the underdamped one in absence of non-conservative forces.  

We illustrated our results using a exactly solvable Brownian heat engine consisting of a Brownian particle 
confined on a one-dimensional ring and subjected to a constant force while in contact with two reservoirs.
By doing so, we showed that the calculations using our overdamped approximation are far simpler than the 
underdamped ones and we also showed that the overdamped long-time heat statistics coincides with the
underdamped one, this time in presence of a non-conservative force. 

We now conclude.
The study of thermal heat engines at the microscale is attracting considerable interest. 
Engines made of Brownian particles are paradigmatic examples of such engines.
Although the underdamped theory can be used to correctly assess the thermodynamic properties of these 
engines, it can be hard to implement in practice because the equations can be very hard to solve.
Since it is often the case in microscale machines that momenta evolve much faster than positions, 
making use of the overdamped approximation is justified and considerably simplifies calculations. 
We thus hope that the overdamped stochastic thermodynamics for multiple reservoirs presented 
in this paper will provide a solid basis for future studies on the performance of Brownian heat engines.

\begin{acknowledgements}
YM appreciate support by the Japan Society for the Promotion of Science through the Program for Leading Graduate Schools (MERIT) and the JSPS Research Fellowship (JSPS KAKENHI Grant Number JP15J00410).
ME was supported by National Research Fund Luxembourg (project FNR/A11/02) and of the European Research Council (project 681456).
This collaboration was done during the overseas dispatch program of Leading Graduate School (MERIT).
YM thanks Masahito Ueda for his insightful comments.
\end{acknowledgements}

\begin{appendix}
\section{Rescaling of variables}
For simplicity, we rescale the quantities as
\eqn{
	x &=& \tilde x\sqrt{\frac{\kb T^{\rm ef}}{m}},\\
	v &=& \tilde v\sqrt{\frac{\kb T^{\rm ef}}{m}},\\
	f &=& \tilde f\sqrt{m\kb T^{\rm ef}},\\
	Q &=& \tilde Q\kb T^{\rm ef},
}
where the tildes indicate that the quantities are rescaled.
Then, the time-evolution equations~(\ref{ULeq1}) and (\ref{ULeq2}) reduce to
\eqn{
	d\tilde x_t &=& \tilde v_tdt,\label{eom1}\\
	d\tilde v_t &=& \tilde f_t(\tilde x_t) dt+
	\sum_\nu \left(
		-\frac{{\tilde \gamma}^\nu}{\epsilon}\tilde  v_t dt + \sqrt{\frac{2{\tilde \gamma}^\nu {\tilde T}^\nu}{\epsilon}} dw^\nu_t
	\right),\label{eom2}\ \ \ \\
	d\tilde Q^\nu_t &=& \frac{1}{\epsilon}{\tilde \gamma}^\nu(\tilde v_t^2-N{\tilde T}^\nu) - \sqrt\frac{2{\tilde \gamma}^\nu {\tilde T}^\nu}{\epsilon} \tilde v_t\cdot dw^\nu_t,\label{eom3}
}
where $\epsilon=m/\gamma^{\rm ef}$ is the overdamped parameter and we define the relative friction coefficient ${\tilde \gamma}^\nu=\gamma^\nu/\gamma^{\rm ef}$ and the relative temperature ${\tilde T}^\nu=T^\nu/T^{\rm ef}$.
We note that these quantities satisfy
\eqn{
	\sum_\nu {\tilde \gamma}^\nu &=& 1,\\
	\sum_\nu {\tilde \gamma}^\nu {\tilde T}^\nu &=& 1.
}
Throughout Appendices, we use these rescaled variables for simplicity.

\section{Derivation of Eq.~(\ref{udteeq})}
Here, we derive the time-evolution equation of the generating function.
This derivation is a generalization of the derivation of the Fokker-Planck equation from the Langevin equation given in Ref.~\cite{Sekimoto}.
We consider a stochastic function
\eqn{
	\mathcal F_t(\tilde x,\tilde v,\{\Lambda^\nu\})=
	\delta(\tilde x_t-\tilde x)
	\delta(\tilde v_t-\tilde v)
	\exp\left[\sum_\nu \Lambda^\nu \tilde Q^\nu_t \right].\ \ \ \ \ 
}
The increment of $\mathcal F_t(\tilde x,\tilde v,\{\Lambda^\nu\})$ can be calculated as
\begin{widetext}
\eqn{
	&&d\mF_t(\tilde x,\tilde v,\{\Lambda^\nu\})\nonumber\\
	&&=
	(\dd_{\tilde x_t} \mF_t) d\tilde x_t
	+ (\dd_{\tilde v_t} \mF_t) d\tilde v_t
	+ \frac{1}{2}(\dd_{\tilde v_t}^2 \mF_t) (d\tilde v_t)^2
	+\sum_\nu \Lambda^\nu \mF_t d\tilde Q^\nu_t
	+ \frac{1}{2}\sum_{\nu,\nu'} \Lambda^\nu\Lambda^{\nu'} \mF_t d\tilde Q^\nu_t d\tilde Q^{\nu'}_t
	+ \sum_\nu \Lambda^\nu (\dd_{\tilde v_t} \mF_t) d{\tilde v_t} dQ^\nu_t\ \ \ \ \ \nonumber\\
	&&=
	-\dd_{\tilde x} (\mF_t d\tilde x_t)
	-\dd_{\tilde v} (\mF_t d\tilde v_t)
	+ \frac{1}{2} \dd_{\tilde v}^2 (\mF_t (d\tilde v_t)^2)
	+\sum_\nu \Lambda^\nu \mF_t d\tilde Q^\nu_t
	+ \frac{1}{2}\sum_{\nu,\nu'}\Lambda^\nu\Lambda^{\nu'} \mF_t d\tilde Q^\nu_t d\tilde Q^{\nu'}_t
	- \sum_\nu \Lambda^\nu \dd_{\tilde v} (\mF_t d{\tilde v_t} d\tilde Q^\nu_t).\label{Fdif}
}
Here, we have to keep the quadratic terms of $d\tilde v_t$ and $d\tilde Q^\nu_t$, since they contain terms proportional to $(dw^\nu_t)^2$, which is the order of $dt$.
To obtain the last line, we use the fact that $\mF$ is an even function with respect to $\tilde x_t-\tilde x$ and $\tilde v_t-\tilde v$.
We average Eq.~(\ref{Fdif}).
For example, the average of the second term on the right-hand side of Eq.~(\ref{Fdif}) is
\eqn{
	\ave{\dd_{\tilde v} (\mF_t d\tilde v_t)}
	&=&
	\dd_{\tilde v} \ave{\mF_t d\tilde v_t}\nonumber\\
	&=&
	\dd_{\tilde v} \ave{\delta(\tilde x_t-\tilde x)
	\delta(\tilde v_t-\tilde v)
	\exp \left[\sum_\nu\Lambda^\nu \tilde Q^\nu_t \right]
	\left[ \tilde f_t(\tilde x_t) dt+
	\sum_\nu \left(
		-\frac{{\tilde \gamma}^\nu}{\epsilon} \tilde v_t dt + \sqrt{\frac{2{\tilde \gamma}^\nu {\tilde T}^\nu}{\epsilon}} dw^\nu_t
	\right)\right]}\nonumber\\
	&=&
	\dd_{\tilde v} \left[
		\left(
			\tilde f_t(\tilde x) dt - \frac{1}{\epsilon} \tilde vdt 
		\right)
			\ave{\delta(\tilde x_t-\tilde x)\delta(\tilde v_t-\tilde v) \exp \left[\sum_\nu\Lambda^\nu \tilde Q^\nu_t \right]}
	\right]\nonumber\\
	&=&
	\tilde f_t(\tilde x) (\dd_{\tilde v} G_t) dt - \frac{1}{\epsilon} \dd_{\tilde v} (\tilde v G_t) dt.
}
The average of the third term on the right-hand side of Eq.~(\ref{Fdif}) is
\eqn{
	\ave{\frac{1}{2} \dd_{\tilde v}^2 (\mF_t (d\tilde v_t)^2)}
	&=&
	\frac{1}{2} \dd_{\tilde v}^2\ave{\mF_t (d\tilde v_t)^2}\nonumber\\		&=&
	\frac{1}{2} \dd_{\tilde v}^2\ave{\delta(\tilde x_t-\tilde x)\delta(\tilde v_t-\tilde v) \exp \left[\sum_\nu\Lambda^\nu \tilde Q^\nu_t \right]
	\left(\sum_\nu \sqrt\frac{2{\tilde \gamma}^\nu{\tilde T}^\nu}{\epsilon} dw^\nu_t \sum_{\nu'} \sqrt\frac{2g^{\nu'}{\tilde T}^{\nu'}}{\epsilon} dw^{\nu'}_t + o(dt)\right)}\nonumber\\
	&=&
	\frac{1}{\epsilon}(\dd_{\tilde v}^2 G_t )dt + o(dt).
}
Other terms in Eq.~(\ref{Fdif}) can be evaluated in a similar way.
As a result, we obtain
\eqn{
	\dd_t G_t =&&
	 -\tilde v\dd_{\tilde x} G_t - \tilde f_t(\tilde x)\dd_{\tilde v} G_t\nonumber\\
	&& + \frac{1}{\epsilon}\left[ \dd_{\tilde v} (\tilde vG_t)
	 +\dd_{\tilde v}^2 G_t
	 +\sum_\nu \Lambda^\nu {\tilde \gamma}^\nu (\tilde v^2-N{\tilde T}^\nu) G_t
	 +\sum_\nu (\Lambda^\nu)^2 {\tilde \gamma}^\nu {\tilde T}^\nu \tilde v^2 G_t
	 +2\sum_\nu \Lambda^\nu {\tilde \gamma}^\nu {\tilde T}^\nu \dd_{\tilde v}(\tilde vG_t)\right],
}
which is equivalent to Eq.~(\ref{udteeq}).
\end{widetext}

\section{Derivation of underdamped fluctuation theorems}
In this Appendix, we derive the three fluctuation theorems~(\ref{udift}), (\ref{uddft}) and (\ref{udssft}) in underdamped dynamics.

\subsection{Finite-time integral fluctuation theorem}
To show the integral fluctuation theorem~(\ref{udift}), we define the initial-point conditioned generating function
\eqn{
	&&G_t(\tilde x,\tilde v,\Lam|\tilde x',\tilde v')\nonumber\\
	&&
	= \ave{
		\delta(\tilde x_t-\tilde x)\delta(\tilde v_t-\tilde v)
			e^{\sum_\nu \Lambda^\nu \tilde Q^\nu_t}
		\Bigl|
		\tilde x_0=\tilde x',\tilde v_0=\tilde v'
	}.\nonumber\\
}
The time-evolution equation of this function is identical to Eq.~(\ref{udteeq}).
Here, we consider the case of $\Lambda^\nu=-1/{\tilde T}^\nu$.
Then, the generating function
$
	G^{\rm r}_t(\tilde x,\tilde v|\tilde x',\tilde v')
	:=G_t(\tilde x,\tilde v,\{-1/{\tilde T}^\nu\}|\tilde x',\tilde v')
$
satisfies
\eqn{\label{teeqGr}
	\dd_t G^{\rm r}_t
	=
	-\tilde v\dd_{\tilde x} G^{\rm r}_t
	- \tilde f_t \dd_{\tilde v} G^{\rm r}_t
	+\frac{1}{\epsilon}\left[
		\dd_{\tilde v}^2 G^{\rm r}_t - \tilde v \dd_{\tilde v} G^{\rm r}_t
	\right].
}
For convenience, we define
\eqn{
	&&G^{\rm t}_t(\tilde x,\tilde v|\tilde x',\tilde v')
	:= \frac{P_t(\tilde x,\tilde v)}{P_0(\tilde x',\tilde v')}
	G^{\rm r}_t(\tilde x,\tilde v|\tilde x',\tilde v')\nonumber\\
	&&=\ave{
		\delta(\tilde x_t-\tilde x)\delta(\tilde v_t-\tilde v)
		e^{
			 -\Delta s_t/\kb- \sum_\nu {\tilde Q^\nu_t}/{{\tilde T}^\nu}
		}
		\Bigl|\tilde x_0=\tilde x',\tilde v_0=\tilde v'
	}.\nonumber\\
}
We note that 
\eqn{\label{}
	G^{\rm t}_0(\tilde x,\tilde v|\tilde x',\tilde v')
	&=&
	\frac{P_0(\tilde x,\tilde v)}{P_0(\tilde x',\tilde v')}
	\delta(\tilde x-\tilde x')\delta(\tilde v-\tilde v')\nonumber\\
	&=&
	\delta(\tilde x-\tilde x')\delta(\tilde v-\tilde v').
}
From Eqs.~(\ref{Keq}) and (\ref{teeqGr}), the time-evolution equation of $G^{\rm t}_t$ is derived as
\eqn{
	\dd_t G^{\rm t}_t
	&=&
	\frac{1}{P_0}\dd_t(G^{\rm r}_t P_t)\nonumber\\
	&=&
	\frac{1}{P_0}((\dd_t G^{\rm r}_t)P_t+G^{\rm r}_t\dd_t P_t)\nonumber\\
	&=&
	-\tilde v \dd_{\tilde x} G^{\rm t}_t
	-\tilde f_t \dd_{\tilde v} G^{\rm t}_t 
	+\frac{1}{\epsilon}[\dd_{\tilde v}^2 G^{\rm t}_t + \dd_{\tilde v}(\tilde vG^{\rm t}_t)]\nonumber\\
	&&\ \ \ \ \ 
	- \frac{2}{\epsilon}
		(\tilde vP_t+\dd_{\tilde v} P_t) \dd_{\tilde v}\left(\frac{G^{\rm t}_t}{P_t}\right).\label{teeqGt}
}
Now, we define
\eqn{\label{defgft}
	G^{\rm t}_t (\tilde x,\tilde v)
	:= \int d\tilde x'd\tilde v'\ 
	G^{\rm t}_t (\tilde x,\tilde v|\tilde x',\tilde v')P_0 (\tilde x',\tilde v').
}
Then, we obtain
\eqn{\label{iniGt}
	G^{\rm t}_0(\tilde x,\tilde v) = P_0(\tilde x,\tilde v),
}
and $G^{\rm t}_t(\tilde x,\tilde v)$ obeys the time-evolution equation~(\ref{teeqGt}).
We can observe that if $G^{\rm t}_t(\tilde x,\tilde v)=P_t(\tilde x,\tilde v)$ the last term on the right-hand side of Eq.~(\ref{teeqGt}) vanishes, and the remaining terms are the ones in the Kramers equation (see Eq.~(\ref{Keq})).
Therefore, together with the initial condition (\ref{iniGt}), we can conclude that the solution of Eq.~(\ref{teeqGt}) is
\eqn{
	G^{\rm t}_t(\tilde x,\tilde v) = P_t(\tilde x,\tilde v).
} 
Therefore, by definition, we have
\eqn{
	\langle
		e^{-\Delta \tilde s^{\rm tot}}
	\rangle
	&=&
	\int dxdv\ G^{\rm t}_\tau(\tilde x,\tilde v)\nonumber\\
	&=&
	\int dxdv\ P_\tau(\tilde x,\tilde v)\nonumber\\
	&=&
	1,
}
which is nothing but the integral fluctuation theorem~(\ref{udift}).

\subsection{Finite-time detailed fluctuation theorem}
Here, we consider the situation in which the system starts from a thermal equilibrium state with the reference reservoir $\nu=0$.
In accordance with the detailed fluctuation theorem~(\ref{uddft}), we consider a new generating function
\eqn{
	\fG_t(\tilde x,\tilde v,\Lambda)
	:=
	\ave{\delta(\tilde x-\tilde x_t)
	\delta(\tilde v-\tilde v_t)
	\exp[-\Lambda {\tilde T}^0 \Delta_{\rm i}\tilde s_t]},
	\nonumber\\
}
where the rescaled irreversible entropy production is defined by
\eqn{
	-{\tilde T}^0 \Delta_{\rm i} \tilde s_t
	:=
	\sum_\nu \eta^\nu \tilde Q^\nu_t - \tilde W_t + \Delta \tilde F_t,
}
where the work and the free energy are rescaled as
$
	F_t = \tilde F_t \kb T^{\rm ef},
	W_t = \tilde W_t \kb T^{\rm ef}.	
$
We can derive the time-evolution equation of $\fG_t$ as
\eqn{
	\dd_t \fG_t
	&=&
	\fL_t(\Lambda) \fG_t
	- \Lambda (\dd_t \tilde V_t)\fG_t
	- \Lambda \tilde f^{\rm nc}_t(\tilde x) \tilde v \fG_t
	+ \Lambda (\dd_t \tilde F_t)\fG_t
	\nonumber\\
	&=&
	\left[
		\fL_t(\Lambda)+{\tilde T}^0\Lambda(\dd_t\ln P^{\rm eq}_t)
		-\Lambda\tilde f^{\rm nc}_t(\tilde x)\tilde v
	\right]
	\fG_t,\label{TEfG}
}
where
$
	P^{\rm eq}_t(\tilde x,\tilde v)
	=
	\exp[{-(\tilde v^2/2+\tilde V_t(\tilde x)-\tilde F)/{\tilde T}^0}]
$
and
$
	\fL_t(\Lambda)
	=
	\mL_t(\{\Lambda^\nu = \eta^\nu \Lambda\})
$.
This time-evolution operator is decomposed as
\eqn{
	\fL_t(\Lambda) &=& \fL^0_t + \frac{1}{\epsilon}\fL^1(\Lambda),\\
	\fL^0_t &=& -\tilde v \dd_{\tilde x} - \tilde f_t \dd_{\tilde v},\\
	\fL^1 (\Lambda)
	&=&
	\dd_{\tilde v}^2 + (1+2\fB)\tilde v\dd_{\tilde v} + (\fA {\tilde v}^2 + N\fB + N),\ \ \ 
}
where
\eqn{
	\fA(\Lambda) &=&
	A(\{\Lambda^\nu = \eta^\nu \Lambda\})\nonumber\\
	&=&
	\Lambda\left(\Lambda-\beta^0\right)
	\left(1-2{\tilde T}^0 + ({\tilde T}^0)^2\sum_\nu\beta^\nu {\tilde \gamma}^\nu\right)\nonumber\\
	&&\ \ \ +\Lambda\left(\beta^0-1\right),\\
	\fB(\Lambda) &=&
	B(\{\Lambda^\nu = \eta^\nu \Lambda\})\nonumber\\
	&=&
	\Lambda(1-{\tilde T}^0),
}
with $\beta^\nu = 1/{\tilde T}^\nu$.
Noting
\eqn{\label{bsymm}
	\fB(\Lambda) + \fB\left(\beta^0-\Lambda\right)
	&=&
	\beta^0-1,\\
	\label{asymm}
	\fA(\Lambda)-\fA\left(\beta^0-\Lambda\right)
	&=&
	\beta^0\left[\fB(\Lambda)
	-\fB\left(\beta^0-\Lambda\right)\right],\ \ \ \ \ 
}
we can confirm the symmetry of
\begin{widetext}
\eqn{\label{fGsymm}
	(P^{\rm eq}_t(\tilde x,\tilde v))^{-1}\fL_t(\Lambda) P^{\rm eq}_t(\tilde x,\tilde v)
	=
	\fL^\dag_{t,\tilde v\to -\tilde v}\left(\beta^0-\Lambda\right)
	+\beta^0\tilde f^{\rm nc}_t(\tilde x)\tilde v,
}
where the subscript $\tilde v\to -\tilde v$ means the velocity inversion.
From Eq.~(\ref{TEfG}), we obtain
\eqn{
	\dd_t((P^{\rm eq}_t)^{-1} \fG_t)
	=
	(P^{\rm eq}_t)^{-1}
	\left[
		 \fL_t(\Lambda)
		+({\tilde T}^0\Lambda - 1)(\dd_t \ln P^{\rm eq}_t)
		-\Lambda\tilde f^{\rm nc}_t(\tilde x) \tilde v
	\right]
	\fG_t.
}
Due to the symmetry~(\ref{fGsymm}), this equation can be rewritten as
\eqn{
	\dd_t((P^{\rm eq}_t)^{-1} \fG_t)
	=
	\left[
		 \fL^\dag_{t,\tilde v\to-\tilde v}\left(\beta^0-\Lambda\right)
		+({\tilde T}^0\Lambda - 1)(\dd_t \ln P^{\rm eq}_t)
		+(\beta^0-\Lambda) \tilde f^{\rm nc}_t(\tilde x)\tilde v
	\right]
	(P^{\rm eq}_t)^{-1} \fG_t.
}
As a result, we obtain
\eqn{
	(P^{\rm eq}_\tau)^{-1}\fG_\tau
	=
	{\bf T} \exp\left[
		\int_0^\tau dt\ \fL^\dag_{t,\tilde v\to-\tilde v}\left(\beta^0-\Lambda\right)
		+({\tilde T}^0\Lambda - 1)(\dd_t \ln P^{\rm eq}_t)
		+(\beta^0-\Lambda)\tilde f^{\rm nc}_t(\tilde x)\tilde v
	\right].
}
When we define the inverted time $\bar t = \tau-t$, this equation can be rewritten as
\eqn{
	\fG_\tau\left(\tilde x,\tilde v,\Lambda\right)
	=
	P^{\rm eq}_{\bar 0}(\tilde x,\tilde v)
	{\bar{\bf T}}
	\exp\left[
		\int_0^\tau d\bar t\ 
		\fL^\dag_{\tau-\bar t,\tilde v\to-\tilde v}\left(\beta^0-\Lambda\right)
		-({\tilde T}^0\Lambda - 1)(\dd_{\bar t} \ln P^{\rm eq}_{\tau-\bar t})
			+(\beta^0-\Lambda)\tilde f^{\rm nc}_{\tau-\bar t}(\tilde x)\tilde v
		\right],
}
where $\bar{\bf T}$ represents the anti-time-ordering operator.
Integrating over $\tilde x$ and $\tilde v$, we obtain
\eqn{\label{FTDT}
	\int d\tilde xd\tilde v\ \fG_\tau\left(\tilde x,\tilde v,\Lambda\right)
	&=&
	\int d\tilde xd\tilde v\ 
	{\bf T}
	\exp\left[
		\int_0^\tau d\bar t\ 
		\fL_{\tau-\bar t,\tilde v\to - \tilde v}\left(\beta^0-\Lambda\right)
		+{\tilde T}^0 (\beta^0-\Lambda)
		(\dd_{\bar t} \ln P^{\rm eq}_{\tau-\bar t})
		+ (\beta^0-\Lambda)\tilde f^{\rm nc}_{\tau-\bar t}(\tilde x)\tilde v
	\right]
	P^{\rm eq}_{\tilde 0}(\tilde x,\tilde v)\nonumber\\
	&=&
	\int d\tilde xd\tilde v\ 
	\bar \fG_{\bar \tau}
	\left(\tilde x,-\tilde v,\beta^0-\Lambda\right),
}
where $\bar \fG$ is the generating function in the reversed process that starts from the equilibrium distribution $P^{\rm eq}_{\bar 0} = P^{\rm eq}_\tau$ and applies the force in the reversed manner.
In terms of the probability distribution function of the irreversible entropy production $P(\Delta_{\rm i} \tilde s)$, Eq.~(\ref{FTDT}) is equivalent to
\eqn{\label{udftdf}
	\int d\Delta_{\rm i} \tilde s\ 
	P(\Delta_{\rm i} \tilde s)
	\exp [-\Lambda {\tilde T}^0 \Delta_{\rm i} \tilde s]
	&=&
	\int d\Delta_{\rm i} \tilde s\ 
	\bar P(\Delta_{\rm i} \tilde s)
	\exp [-(\beta^0-\Lambda){\tilde T}^0 \Delta_{\rm i} \tilde s]
	\nonumber\\
	&=&
	\int d\Delta_{\rm i} \tilde s\ 
	\bar P(-\Delta_{\rm i} \tilde s)e^{\Delta_{\rm i} \tilde s}
	\exp [-\Lambda{\tilde T}^0\Delta_{\rm i} \tilde s],
}
\end{widetext}
where $\bar P$ represents the probability in the reversed process.
Equation~(\ref{udftdf}) indicates the finite-time detailed fluctuation theorem of the irreversible entropy production,
namely,
\eqn{
	\frac{\bar P(-\Delta_{\rm i} \tilde s)}
	{P(\Delta_{\rm i} \tilde s)}
	=
	\exp\left[
		-\Delta_{\rm i} \tilde s
	\right],
}
which is Eq.~(\ref{uddft}) in the original scale.

\subsection{Asymptotic steady-state fluctuation theorem}
Here, we consider the fluctuation symmetry of the generator $\mL_t(\Lam)$.
We note that
\eqn{
	A\left(\left\{
		-\Lambda^\nu -\beta^\nu
	\right\}\right)
	&=&
	A(\Lam),\label{symA}\\
	B\left(\left\{
		-\Lambda^\nu -\beta^\nu
	\right\}\right)
	&=&
	-B(\Lam)-1.\label{symB}
}
As a result, we obtain
\eqn{
	\mL^{1\dag}\left(\left\{
		-\Lambda^\nu -{\beta^\nu}
	\right\}\right)
	=
	\mL^1(\Lam),
}
where the superscript $\dag$ represents the adjoint operator.
On the other hand, we have
\eqn{
	\mL^{0\dag}_t=v\dd_x + f_t\dd_v = \mL^0_{t,v\to -v},
} 
where the subscript $v\to-v$ means the velocity inversion.
Hence, we have
\eqn{
	\mL^\dag_t \left(\left\{
		-\Lambda^\nu -\beta^\nu
	\right\}\right)
	=
	\mL_{t,v\to -v} (\Lam),
}
which is nothing but Eq.~(\ref{udssft}).

Since the adjoint operation and the velocity inversion do not change eigenvalues of a differential operator, we conclude that the eigenvalues of $\mL_t(\Lam)$ are the same as those of $\mL_t (\{-\Lambda^\nu-\beta^\nu\})$.
Consequently, when the force $f_t(x)$ is time-independent, the generating function asymptotically has the symmetry
\eqn{\label{SSFT}
	\lim_{t\to \infty} \frac{1}{t}\ln\frac{\mG_t(\Lam)}{\mG_t (\{-\Lambda^\nu - \beta^\nu\})}
	=
	0,
}
because the long-time behavior is dominated by the largest eigenvalue of $\mL(\Lam)$.
One might therefore expect that the asymptotic steady-state fluctuation theorem for the joint probability distribution of heats holds under time-independent force.
However, this is not the case due to non-analyticity.

For ordinary situations, the large deviation function of the probability distribution function $h(\{q^\nu\})$ defined by
\eqn{
	P(\{Q^\nu_t\})\simeq \exp\left[
		t h(\{Q^\nu_t/t\})
	\right]
}
is related to the largest eigenvalue $\alpha(\Lam)$ of $\mL(\Lam)$ by the Legendre transform as
\eqn{\label{Leg}
	h(\{q^\nu\})
	=
	\min_{\{\Lambda^\nu\}}\left[
		\alpha(\{\Lambda^\nu\}) - \sum_\nu q^\nu \Lambda^\nu
	\right].
}
As a result, the symmetry of the generating function~(\ref{SSFT}) leads to the fluctuation theorem
\eqn{\label{appudssft}
	\lim_{t\to\infty} \frac{1}{t}\ln
	\frac{P(\{\tilde Q^\nu_t=\tilde q^\nu t\})}{P(\{\tilde Q^\nu_t=-\tilde q^\nu t\})}
	=
	\sum_\nu \frac{\tilde q^\nu}{{\tilde T}^\nu}.
}
Equation~(\ref{Leg}) is guaranteed by the saddle-point approximation in the long-time limit.
However, when the asymptotic behavior of the generating function
\eqn{
	\mG_t(\Lam) \simeq g(\Lam) \exp [t \alpha(\Lam)]
}
has a singularity in that $g(\Lam)$ has a pole in the region of the saddle-point approximation, Eq.~(\ref{Leg}) does not hold.
Consequently, the fluctuation theorem~(\ref{appudssft}) ceases to be valid \cite{vZC03,vZC04,Vis06,Sab12,NP12}, although the symmetry of the time-evolution operator~(\ref{udssft}) holds.
Some observations imply that this is true for our case and therefore the steady-state fluctuation theorem~(\ref{appudssft}) would not hold.
Clarifying this point is left as a topic for future research.

\section{Derivation of overdamped approximation}
In this Appendix, we derive overdamped time-evolution equation from Eqs.~(\ref{eq-1}-\ref{eq1}).

First of all, we search for the eigenvalues of $\mL^1(\Lam)$.
The equation for the right eigenfunctions can be written as
$
	\mL^1(\Lam) \phi(\tilde v,\Lam)
	=
	\alpha(\Lam) \phi(\tilde v,\Lam),
$
or equivalently
\eqn{
	\dd_{\tilde v}^2 \phi + (1+2B) {\tilde v}\dd_{\tilde v} \phi + (A{\tilde v}^2+NB+N-\alpha)\phi
	=
	0.\ \ \ \ \ 
}
To simplify this equation, we transform $\phi$ as
\eqn{
	\phi (\tilde v,\Lam)
	=
	\varphi (\tilde v, \Lam)
	\exp\left[
		-\frac{1}{2} \kappa(\Lam) {\tilde v}^2
	\right],
}
where $\kappa$ is a yet undetermined constant.
Then, we obtain
\begin{widetext}
\eqn{
	\dd_{\tilde v}^2 \varphi
	+ (1+2B-2\kappa) {\tilde v}\dd_{\tilde v}\varphi
	+\left[
		(\kappa^2-(1+2B)\kappa+A){\tilde v}^2+N(1+B-\kappa)-\alpha
	\right]\varphi=0.
}
\end{widetext}
To eliminate the coefficient of ${\tilde v}^2\varphi$, we set
$
	\kappa = ({1+2B+R})/{2},
$
where
$
	R=\sqrt{(1+2B)^2-4A}.
$
Then, we acquire
\eqn{
	\dd_{\tilde v}^2\varphi
	-R{\tilde v}\dd_{\tilde v}\varphi
	+\left(\frac{N(1-R)}{2}-\alpha\right)\varphi
	= 0.
}
If we define a new variable as
$
	u=\sqrt{{R}/{2}} \tilde v,
$
the equation reduces to
\eqn{
	\dd_u^2\varphi - 2u\dd_u\varphi + \frac{N(1-R)-2\alpha}{R}\varphi=0,
}
which is of the form of the Hermite equation.
Therefore, the eigenfunctions, which vanish at the infinity, are
\eqn{
	\varphi_{\{n_i\}} = \prod_{i=1}^N H_{n_i}(u_i),
}
where $u_i$ is the $i$-th component of $u$ and $H_n$ is the Hermite polynomial of the $n$-th degree, and $n$ is an integer larger than or equal to zero.
The eigenvalues satisfy
\eqn{
	2\sum_{i=1}^N n_i &=&\frac{N(1-R)-2\alpha_{\{n_i\}}}{R},\nonumber\\
}

In summary, $\mL^1$ has the eigenvalues
\eqn{
	\alpha_{\{n_i\}}=\frac{N(1-R)}{2}-R\sum_{i=1}^N n_i,
}
and right eigenfunctions
\eqn{
	\phi_{\{n_i\}} = \prod_{i=1}^N H_{n_i}\left(\sqrt\frac{R}{2} \tilde v_i\right)
	\exp\left[
		-\frac{1}{2} \kappa \tilde v^2
	\right].
}
In a similar way, we can confirm that the left eigenfunctions are
\eqn{
	\bar\phi_{\{n_i\}}
	=
	\prod_{i=1}^N H_{n_i}\left(\sqrt\frac{R}{2} \tilde v_i\right)
	\exp\left[
		-\frac{1}{2} \rho \tilde v^2
	\right],
}
where
$
	\rho = ({-1-2B+R})/{2}.
$

From Eq.~(\ref{eq-1}), we observe that the eigenfunction $\phi_0$ corresponding to the largest eigenvalue $\alpha_0$ survives after relaxation in the fast timescale, and the other eigenfunctions vanish.
Hence, the leading order of the generating function can be written as
\eqn{\label{G0}
	&&G^{(0)}_{\theta,t,{\hat t}}(\tilde x,\tilde v,\Lam)=
	\hat G^{(0)}_{t,{\hat t}}(\tilde x,\Lam)
	\phi_0(\tilde v) e^{\alpha_0 \theta},
}
where $\hat G^{(0)}_{t,\hat t}(\tilde x,\Lam)$ is an arbitrary function independent from $\theta$ and $\tilde v$.
Then, Eq.~(\ref{eq0}) reduces to
\eqn{
	&&(\dd_\theta-\mL^1)G^{(1)}_{\theta,t,{\hat t}}\nonumber\\
	&&= -(\dd_t - \mL^0_{t,{\hat t}}) G^{(0)}_{\theta,t,{\hat t}}\nonumber\\
	&&=
	-\left(\dd_t \hat G^{(0)}_{t,{\hat t}}
	+\tilde v\dd_{\tilde x} \hat G^{(0)}_{t,{\hat t}}
	-\kappa \tilde f_{t,{\hat t}} \tilde v \hat G^{(0)}_{t,{\hat t}}\right)
	\phi_0 e^{\alpha_0 \theta}.
}
The $\theta$-dependence of the right-hand side is $e^{\alpha_0\theta}$, and therefore the left-hand side should have the same dependence.
Hence, we can replace $\dd_\theta$ in the left-hand side by $\alpha_0$, and obtain
\eqn{\label{eq0'}
	&&\left(
		 \alpha_0 - \mL^1
	\right)G^{(1)}_{\theta,t,{\hat t}}\nonumber\\
	&&=
	-\left(\dd_t \tilde G^{(0)}_{t,{\hat t}}
	+\tilde v\dd_{\tilde x} \tilde G^{(0)}_{t,{\hat t}}
	-\kappa \tilde f_{t,{\hat t}} \tilde v \tilde G^{(0)}_{t,{\hat t}}\right)
	\phi_0 e^{\alpha_0 \theta}.
}
We note that the assumption that the force is independent of $\theta$ is crucial in this step.
Multiplying $\bar \phi_0$ from left and integrating over $v$, we obtain
$
	0=\dd_t \tilde G^{(0)}_{t,{\hat t}}.
$
Then, noting
$
	\mL^1(\tilde v\phi_0) = (\alpha_0-NR)\tilde v\phi_0,
$
we can explicitly solve Eq.~(\ref{eq0'}) as
\eqn{\label{G1}
	&&G^{(1)}_{\theta,t,{\hat t}}(\tilde x,\tilde v,\Lam)\nonumber\\
	&&=
	\left[
		\hat G^{(1)}_{t,{\hat t}}(\tilde x,\Lam)
		- \frac{1}{NR}\tilde v (\dd_{\tilde x}-\kappa \tilde f_{t,{\hat t}})
		\hat G^{(0)}_{{\hat t}}(\tilde x,\Lam)
	\right]
	\phi_0 e^{\alpha_0\theta},\nonumber\\
}
where $\hat G^{(1)}_{t,{\hat t}}(x,\Lam)$ is an arbitrary function.

The next order equation~(\ref{eq1}) reads
\eqn{
	(\dd_\theta-\mL^1) G^{(2)}_{\theta,t,{\hat t}}
	=
	-(\dd_t-\mL^0_{t,{\hat t}})G^{(1)}_{\theta,t,{\hat t}}
	- \dd_{{\hat t}} G^{(0)}_{\theta,t,{\hat t}}.
}
The only $\theta$ dependence is $e^{\alpha_0\theta}$, and therefore we obtain
\eqn{
	(\alpha_0-\mL^1) G^{(2)}_{\theta,t,{\hat t}}
	=
	-(\dd_t-\mL^0_{t,{\hat t}})G^{(1)}_{\theta,t,{\hat t}}
	- \dd_{{\hat t}} G^{(0)}_{\theta,t,{\hat t}}.
}
Multiplying $\tilde \phi_0$ from left and integrating over $\tilde v$, we obtain
\eqn{
	0&=&
	-\dd_t \hat G^{(1)}_{t,{\hat t}}
	+ \frac{1}{R^2}
	\dd_{\tilde x}(\dd_{\tilde x}-\kappa \tilde f_{t,{\hat t}})\hat G^{(0)}_{{\hat t}}
	\nonumber\\
	&&
	+\frac{R-\kappa}{R^2}\tilde f_{t,{\hat t}}(\dd_{\tilde x}-\kappa\tilde f_{t,\hat t} )\hat G^{(0)}_{\hat t}
	-\dd_{{\hat t}} \hat G^{(0)}_{{\hat t}},
}
and equivalently
\eqn{
	&&\dd_{{\hat t}} \hat G^{(0)}_{{\hat t}} + \dd_t \hat G^{(1)}_{t,{\hat t}}\nonumber\\
	&&=\frac{1}{R^2}
	[\dd_{\tilde x}^2
	-\kappa\dd_{\tilde x}(\tilde f_{t,{\hat t}}\cdot)
	+\rho \tilde f_{t,{\hat t}}\dd_{\tilde x}
	+A \tilde f^2_{t,{\hat t}}]\hat G^{(0)}_{{\hat t}}.
}
Therefore, when we define
\eqn{
	&&G\od_t(\tilde x,\Lam)\nonumber\\
	&&=
	\left(\frac{2\pi}{\kappa}\right)^{N/2} 
	\left[\hat G^{(0)}_{{\hat t} = \epsilon t}(\tilde x,\Lam)
	+\epsilon \hat G^{(1)}_t(\tilde x,\Lam)\right],
}
the time-evolution equation reduces to
\eqn{
	\dd_t G\od_t
	=
	\tilde\mL\od_t(\Lam) G\od_t+\mathcal O(\epsilon^2),
}
where
\eqn{
	\tilde\mL\od_t(\Lam)
	=
	\frac{\epsilon}{R^2}[
		\dd_{\tilde x}^2-\kappa\dd_{\tilde x}(\tilde f_t\cdot)
		+\rho \tilde f_t\dd_{\tilde x} + A \tilde f^2_t
	].\nonumber\\
}

In terms of $G\od_t$, the underdamped generating function $G_t$ is written as
\eqn{
	G_t
	&=&
	\left(\frac{\kappa}{2\pi}\right)^{\frac N 2}
	\left[
		1-\frac{\epsilon}{NR}\tilde v(\dd_{\tilde x}-\kappa \tilde f)+\mathcal O (\epsilon^2)
	\right]
	G\od_t \phi_0 e^{\alpha_0 \theta}.\nonumber\\
}
Integrating over $\tilde v$, we obtain
\eqn{\label{Gv}
	\int d\tilde v\ G_t
	=
	\left[
		G\od_t + \mathcal O (\epsilon^2)
	\right]
	\exp\left[
		\frac{\alpha_0 t}{\epsilon}
	\right].
}
Setting $t=0$, we obtain the initial condition
$
	G\od_0(\tilde x) = P\od_0(\tilde x),
$
where
$
	P\od_t(\tilde x) = \int d\tilde v\ P_t(\tilde x,\tilde v).
$
Integrating Eq.~(\ref{Gv}) over $\tilde x$, we obtain
\eqn{\label{GGod}
	\mG_t
	=
	[\mG\od_t+{\mathcal O}(\epsilon^2)]
	\exp\left[
		\frac{\alpha_0 t}{\epsilon}
	\right],
}
where we define the overdamped generating function as
\eqn{
	\mG\od_t(\Lam) = \int d\tilde x\ G\od_t(\tilde x,\Lam).
}

\section{Time evolution of the generating function derived from the overdamped Langevin equation}
In this section, we consider the time evolution of the generating function in the overdamped system coupled to a single heat reservoir.
After the rescaling, the Langevin equation reads
\eqn{
	d\tilde x_t
	=
	\frac{1}{\epsilon} \tilde f_t dt + \sqrt\frac{2}{\epsilon} dw_t.
}
The heat generation is written as
\eqn{
	d\tilde Q_t
	&=&
	\tilde f_t\circ\dot {\tilde x}_t dt\nonumber\\
	&=&
	\frac{1}{\epsilon}f_t^2 dt + \sqrt\frac{2}{\epsilon}f_t\circ dw_t.
}
We define the generating
\eqn{
	G^{\rm od,sing}_t(\tilde x,\Lambda)
	:=
	\ave{\delta(\tilde x-\tilde x_t)e^{\Lambda \tilde Q_t}},
}
then the time-evolution can be derived as
\eqn{
	\dd_t G^{\rm od,sing}_t
	=
	\tilde \mL^{\rm od,sing}_t(\Lambda) G^{\rm od,sing}_t,
}
where 
\eqn{\label{odsing}
	&&\tilde \mL^{\rm od,sing}_t(\Lambda)\nonumber\\
	&&=
	\frac{1}{\epsilon}
	[\dd_{\tilde x}^2-(1+\Lambda)\dd_{\tilde x}(\tilde f_t\cdot)
	-\Lambda \tilde f_t\dd_{\tilde x} + \Lambda(1+\Lambda)\tilde f_t^2].
}
When ${\tilde T}^\nu=1,\ \Lambda^\nu=\Lambda$ for all $\nu$, Eq.~(\ref{odteop}) reduces to Eq.~(\ref{odsing}).
We note that we cannot derive a similar result from the overdamped Langevin equation in presence of two or more heat reservoirs, because the heats themselves are ill-defined as discussed in the main text.

\section{Derivation of overdamped fluctuation theorems}
In this Appendix, we derive three overdamped fluctuation theorems, which correspond to the three underdamped fluctuation theorems derived in Appendix C.

\subsection{Finite-time integral fluctuation theorem}
We define a reduced generating function by
\eqn{
	G^{\rm od,r}_t(\tilde x|\tilde x')
	:=
	G^{\rm od}_t (\tilde x,\{\Lambda^\nu=-1/\tilde T^\nu\}|\tilde x'),
}
which counts the entropy production in the heat reservoirs.
The time-evolution equation is given by
\eqn{\label{odteGb}
	\frac{1}{\epsilon} \dd_t G^{\rm od, r}_t
	&=&
	\tilde\mL^{\rm od}_t(\{\Lambda^\nu=-1/\tilde T^\nu\}) G^{\rm od, r}_t
	\nonumber\\
	&=&
	\dd_{\tilde x}^2 G^{\rm od,r}_t + \tilde f_t \dd_{\tilde x} G^{\rm od,r}_t
}
On the other hand, the marginal probability distribution $P\od_t(\tilde x)$ satisfies the Fokker-Planck equation
\eqn{\label{FPeqn}
	\frac{1}{\epsilon} \dd_t P\od_t
	=
	\dd_{\tilde x}^2 P\od_t - \dd_{\tilde x}(\tilde f_t P\od_t).
}
Let us define
\eqn{
	G^{\rm od,t}_t(\tilde x|\tilde x')
	:=
	\frac{P\od_t(\tilde x)}{P\od_0(\tilde x')}
	G^{\rm od,r}_t(\tilde x|\tilde x'),
}
then, from Eqs.~(\ref{odteGb}) and (\ref{FPeqn}), the time evolution is given by
\eqn{\label{odteGt}
	\frac{1}{\epsilon}\dd_t G^{\rm od,t}_t
	&=&
	\dd_{\tilde x}^2 G^{\rm od,t}_t - \dd_{\tilde x}(\tilde f_tG^{\rm od,t}_t)\nonumber\\
	&&\ \ \ 
	+2(\tilde f_t P\od_t-\dd_{\tilde x} P\od_t) \dd_{\tilde x}
	\left(
		\frac{G^{\rm od,t}_t}{P\od_t}
	\right).
}
The unconditioned generating function
\eqn{
	G^{\rm od,t}_t(\tilde x)
	=
	\int d\tilde x'\ G^{\rm od,t}_t(\tilde x|\tilde x') P\od_0(\tilde x')
}
satisfies the same time-evolution equation, and the initial condition
is
$
	G^{\rm od,t}_0(\tilde x) = P\od_0(\tilde x).
$
Since the last term on the right-hand side of Eq.~(\ref{odteGt}) vanishes in the case of $G^{\rm od,t}_t(\tilde x)=P_t\od(\tilde x)$, we can conclude that
\eqn{
	G^{\rm od,t}_t(\tilde x) = P\od_t(\tilde x).
}
From the normalization of the probability distribution function, we obtain
\eqn{
	\int d\tilde x\ G^{\rm od,t}_t(\tilde x) = 1,
}
which is the integral fluctuation theorem~(\ref{odift}).

\subsection{Finite-time detailed fluctuation theorem}
Next, we consider the finite-time detailed fluctuation theorem.
To this aim, the non-conservative work should be defined in the overdamped approximation and therefore we have to generalize the approximation.
We start from the underdamped generating function
\eqn{
	&&\check G_t(\tilde x,\tilde v,\Lam, \lambda)\nonumber\\
	&&=
	\left\langle
		\delta (\tilde x_t-\tilde x) \delta (\tilde v_t-\tilde v)
		\exp\left[
			\sum_\nu \Lambda^\nu \tilde Q^\nu_t + \lambda \tilde W^{\rm nc}_t
		\right]
	\right\rangle,
	\ \ \ \ \ 
}
which counts the non-conservative work as well as the heat flows.
By going through almost the same procedure, we obtain the overdamped time-evolution operator
\eqn{
	\mathcal{\check L}^{\rm od}_t(\Lam,\lambda)
	&=&
	\frac{\epsilon}{R^2}\huge[
		\partial_{\tilde x}^2
		- \partial_{\tilde x} [(\kappa \tilde f_t + \lambda \tilde f_t^{\rm nc})\cdot]
		+(\rho \tilde f_t -\lambda \tilde f_t^{\rm nc})\partial_{\tilde x}
		\nonumber\\
		&&\ \ \ \ \ \ \ \ \ \ 
		-(\rho \tilde f_t -\lambda \tilde f_t^{\rm nc})(\kappa \tilde f_t + \lambda \tilde f_t^{\rm nc})
	\huge]
}
for the overdamped generating function
\eqn{
	&&\check G_t^{\rm od}(\tilde x,\Lam,\lambda)\nonumber\\
	&&=
	\left\langle
		\delta(\tilde x-\tilde x_t)
		\exp\left[
			\sum_\nu \Lambda^\nu Q^{\nu, \rm od}_t + \lambda W^{\rm nc,od}
		\right]
	\right\rangle.
}
For convenience, we split the time-evolution operator as
\eqn{\label{spL}
	\mathcal{\check L}^{\rm od}_t(\Lam,\lambda)
	=
	\mathcal{L}^{\rm c}_t(\Lam)
	+
	\mathcal{L}^{\rm nc}_t(\Lam,\lambda),\ \ \ \ \ 
}
where the first term is the time-evolution operator~(\ref{odteop}) for vanishing non-conservative force
\eqn{
	\mathcal{L}^{\rm c}_t(\Lam)
	:=
	\mathcal{L}^{\rm od}_t(\Lam)|_{\tilde f^{\rm nc}=0}
}
 and the remaining terms are packed into the second term
\eqn{
	&&\mL^{\rm nc}_t(\Lam,\lambda)
	\nonumber\\
	&&
	:=
	\frac{\epsilon}{R^2}[
		-(\lambda+\kappa)\dd_{\tilde x}(\tilde f_t^{\rm nc}\cdot)
		-(\lambda-\rho)\tilde f^{\rm nc}_t \dd_{\tilde x}
		\nonumber\\
		&&\ \ \ \ \ 
		-(\rho\lambda-\kappa\lambda+2\kappa\rho)(\dd_{\tilde x} V_t)\tilde f^{\rm nc}_t
		+(\rho-\lambda)(\kappa+\lambda)(\tilde f^{\rm nc})^2
	].
	\nonumber\\
}

With this generalization, we derive the detailed fluctuation theorem.
We consider a situation in which the system starts from an equilibrium state with respect to the reference reservoir $\nu=0$.
We define
\eqn{
	\fG\od_t(\tilde x, \Lambda)
	:=
	\langle \delta (\tilde x - \tilde x_t)
	\exp[-\Lambda\tilde T^0 \Delta_{\rm i}\tilde s\od]
	\rangle,
}
where
\eqn{
	-\tilde T^0 \Delta_{\rm i}\tilde s\od
	=
	\Delta \tilde F\od - \tilde W + \sum_\nu \eta^\nu \tilde Q^{\rm od,\nu}.
}
The time evolution of $\fG\od_t$ is given by
\eqn{
	\dd_t \fG\od_t
	&=&
	\check\fL\od_t(\Lambda)\fG\od_t
	+ \Lambda(\dd_t \tilde F\od_t)\fG\od_t
	- \Lambda(\dd_t \tilde V_t)\fG\od_t
	\nonumber\\
	&=&
	[\check\fL\od_t(\Lambda)
	+ \tilde T^0 \Lambda (\dd_t \ln P^{\rm od, eq}_t)]
	\fG\od_t,
}
where we define the time-evolution operator by
\eqn{
	\check\fL^{\rm od}_t(\Lambda)
	&:=&
	\check\mL^{\rm od}_t(\{\Lambda^\nu=\eta^\nu\Lambda\},\lambda=-\Lambda).
}
In accordance with Eq.~(\ref{spL}), $\check\fL^{\rm od}_t(\Lambda)$ is split as
\eqn{
	\check\fL^{\rm od}_t(\Lambda)
	=
	\fL_t^{\rm c}(\Lambda) + \fL_t^{\rm nc}(\Lambda),
}
where the conservative and non-conservative parts are defined by
\eqn{
	\fL_t^{\rm c}(\Lambda)
	&=&
	\mL^{\rm c}_t(\{\Lambda^\nu=\eta^\nu\Lambda\}),\\
	\fL_t^{\rm nc}(\Lambda)
	&=&
	\mL_t^{\rm nc}(\{\Lambda^\nu=\eta^\nu\Lambda\},\lambda=-\Lambda).
}

The explicit form of the conservative part reads
\eqn{
	\fL^{\rm c}_t(\Lambda)
	&=&
	\frac{\epsilon}{\fR^2}
	[
		\dd_x^2
		+ \mathfrak{K}\dd_x((\dd_x \tilde V_t)\cdot)
		\nonumber\\
		&&\ \ \ \ \ 
		-\mathfrak{P}(\dd_x \tilde V_t)\dd_x
		+ \fA (\dd_x \tilde V_t)^2
	],
}
where $\fR(\Lambda) = R(\{\Lambda^\nu=\eta^\nu\Lambda\})$,
$\mathfrak{K}(\Lambda) = \kappa(\{\Lambda^\nu=\eta^\nu\Lambda\})$ and $\mathfrak{P}(\Lambda) = \rho(\{\Lambda^\nu=\eta^\nu\Lambda\})$.
By simple calculation, we obtain
\eqn{
	\fR\left(\beta^0-\Lambda\right)
	&=&
	\fR(\Lambda),\\
	\fK\left(\beta^0-\Lambda\right)
	&=&
	\fP(\Lambda)+\beta^0,\\
	\fP\left(\beta^0-\Lambda\right)
	&=&
	\fK(\Lambda)-\beta^0.
}
Consequently, we can obtain the symmetry of
\eqn{
	(P^{\rm od,eq}_t(\tilde x))^{-1}
	\fL^{\rm c}_t(\Lambda)
	P^{\rm od,eq}_t(\tilde x)
	=
	\fL^{\rm c\dag}_t(\beta^0-\Lambda).
}
In a similar manner, the same symmetry can be shown for the non-conservative part
\eqn{
	(P^{\rm od,eq}_t(\tilde x))^{-1}
	\fL^{\rm nc}_t(\Lambda)
	P^{\rm od,eq}_t(\tilde x)
	=
	\fL^{\rm nc\dag}_t(\beta^0-\Lambda).
}
Therefore, the sum of them has the same symmetry
\eqn{
	(P^{\rm od,eq}_t(\tilde x))^{-1}
	\check\fL^{\rm od}_t(\Lambda)
	P^{\rm od,eq}_t(\tilde x)
	=
	\check\fL^{\rm od\dag}_t(\beta^0-\Lambda).
}
As a result, a similar procedure to the one in Appendix~C~2 leads to the symmetry of the overdamped generating function as
\eqn{
	\int d\tilde x\ \fG\od_\tau(\tilde x,\Lambda)
	=
	\int d\tilde x\ \bar\fG\od_{\bar \tau}(\tilde x,\beta^0-\Lambda),
}
which means the detailed fluctuation theorem~(\ref{oddft}).

\subsection{Asymptotic steady-state fluctuation theorem}
Noting Eqs.~(\ref{symA}) and (\ref{symB}), we obtain
\eqn{
	R\left(\left\{
		-\Lambda^\nu-\beta^\nu
	\right\}\right)
	&=&
	R(\Lam),\\
	\kappa\left(\left\{
		-\Lambda^\nu-\beta^\nu
	\right\}\right)
	&=&
	\rho(\Lam),\\
	\rho\left(\left\{
		-\Lambda^\nu-\beta^\nu
	\right\}\right)
	&=&
	\kappa(\Lam).
}
Therefore, $\mL\od$ has the symmetry
\eqn{
	\mL\od\left(\left\{
		-\Lambda^\nu-\beta^\nu
	\right\}\right)
	=
	\mL\od{}^\dag(\Lam).
}

\section{Details of underdamped calculation}
In terms of the rescaled variables in Appendix A, Eq.~(\ref{exteop}) reads
\eqn{\label{rexteop}
	\mL_v = -\tilde f\dd_{\tilde v} + \frac{1}{\epsilon}[\dd_{\tilde v}^2 + (1+2B)\dd_{\tilde v}(\tilde v\cdot)+A{\tilde v}^2-B].\ \ \ 
}
The initial condition is rewritten as
\eqn{
	G_0(\tilde v,\Lam)
	=
	\frac{1}{\sqrt{2\pi}}
	\exp\left[
		-\frac{1}{2}(\tilde v-\epsilon \tilde f)^2
	\right].
}
By a method similar to that in Appendix B, we can obtain the eigenvalues of the operator (\ref{rexteop})
\eqn{
	\alpha_n := \frac{1-R}{2}+\frac{\epsilon^2 A \tilde f^2}{R^2} - Rn
}
and the corresponding right eigenfunctions
\eqn{
	\psi_n(\tilde v)
	=
	&&H_n (\tilde u)
	\exp\left[
		-\frac{1}{2}\kappa \left(\tilde v-\frac{\epsilon \tilde f}{R}\right)^2
	\right],
}
where
\eqn{
	\tilde u=\sqrt\frac{R}{2}
		\left(
			\tilde v - \frac{\kappa-\rho}{R^2}\epsilon \tilde f
		\right).
}
Therefore, the generating function can be expanded as
\eqn{
	G(\tilde v,\Lam)
	=
	\sum_{n=0}^\infty
	c_n \psi_n(\tilde v)\exp\left[
		\frac{1}{\epsilon}(\alpha_0-Rn)t
	\right].
}
The initial condition is
\eqn{
	\frac{1}{\sqrt{2\pi}}
	\exp\left[
		-\frac{1}{2}(\tilde v-\epsilon \tilde f)^2
	\right]
	=
	\sum_{n=0}^\infty
	c_n \psi_n(\tilde v),
}
or equivalently
\begin{widetext}
\eqn{
	\sum_{n=0}^\infty 
	c_n H_n \left(
		\tilde u
	\right)
	\exp\left[
		-\tilde u^2
	\right]
	=
	\exp\left[
		-\frac{1}{2}(\tilde v-\epsilon \tilde f)^2
		+ \frac{1}{2}\kappa \left(\tilde v-\frac{\epsilon \tilde f}{R}\right)^2
		-\tilde u^2
	\right].
}
Hence, from the orthogonality of the Hermite polynomials, we obtain
\eqn{
	c_n = \frac{1}{\sqrt{2}\pi 2^n n!}
	\int d\tilde u\ H_n(\tilde u)
	\exp\left[
		-\frac{1}{2}(\tilde v-\epsilon \tilde f)^2
		+ \frac{1}{2}\kappa \left(\tilde v-\frac{\epsilon \tilde f}{R}\right)^2
		-\tilde u^2
	\right].\ \ \ 
}
Then, the generating function is explicitly written as
\eqn{
	G_t(\tilde v,\Lam)
	=&&
	\sum_{n=0}^\infty
	\frac{1}{\sqrt{2}\pi 2^n n!}
	H_n(\tilde u) \exp\left[
		-\frac{\kappa}{2}\left(\tilde v-\frac{\epsilon \tilde f}{R}\right)^2
		+\frac{1}{\epsilon}(\alpha_0-Rn)t
	\right]\nonumber\\
	&&\ \ \ \times
	\int d\tilde u'\ H_n(\tilde u')
	\exp\left[
		-\frac{1}{2}(\tilde v'-\epsilon \tilde f)^2
		+ \frac{\kappa}{2} \left(\tilde v'-\frac{\epsilon \tilde f}{R}\right)^2
		-\tilde u'^2
	\right].
}
By using a formula
\eqn{
	\sum_{n=0}^\infty \frac{x^n}{2^n n!} H_n(\tilde u) H_n(\tilde u')
	=
	\frac{1}{\sqrt{1-x^2}}\exp\left[
		\frac{2x\tilde u\tilde u'-(\tilde u^2+\tilde u'^2)x^2}{1-x^2}
	\right],
}
we obtain
\eqn{
	G_t(\tilde v,\Lam)
	=&&
	\int d\tilde u'\ 
	\frac{1}{\sqrt 2 \pi \sqrt{1-e^{-2Rt/\epsilon}}}\nonumber\\
	&&\ \ \ \times\exp\left[
		\frac{2e^{-Rt/\epsilon}\tilde u \tilde u'
		-e^{-2Rt/\epsilon}(\tilde u^2+\tilde u'^2)}
		{1-e^{-2Rt}}
		-\frac{\kappa}{2}\left(\tilde v-\frac{\epsilon \tilde f}{R}\right)^2
		+\frac{\alpha_0 t}{\epsilon}
		-\frac{1}{2}(\tilde v'-\epsilon \tilde f)^2
		+ \frac{\kappa}{2} \left(\tilde v'-\frac{\epsilon \tilde f}{R}\right)^2
		-\tilde u'^2
	\right].\nonumber\\
}
\end{widetext}
We can analytically conduct the integration of $\tilde u'$ because the integrand is a Gaussian with respect to $\tilde u'$, but the result is too long to show here.
As a result, we can observe that $G_t(\tilde v,\Lam)$ is also a Gaussian with respect to $\tilde v$.
Therefore, we can obtain $\mG_t(\Lam)=\int d\tilde v\ G_t(\tilde v, \Lam)$ and $\mC_t(\Lam)=\ln \mG_t(\Lam)$ analytically, which are also too long to show here.

\end{appendix}

\bibliography{reference}

\end{document}